\documentclass{jfm}

\usepackage{graphicx}
\usepackage{newtxtext}
\usepackage{newtxmath}
\usepackage{natbib}
\usepackage{hyperref}
\usepackage{caption3}
\usepackage{caption} 
\usepackage{subcaption}
\usepackage[dvipsnames]{xcolor}
\usepackage{soul}
\setstcolor{blue}
\usepackage{contour}
\usepackage{color}

\usepackage{amsmath}
\usepackage{amssymb}
\usepackage{natbib}

\usepackage{graphicx}
\usepackage{dcolumn}
\usepackage{bm}
\usepackage{comment}
\usepackage{float} 

\newcommand{\revise}[1]{{\color{black}#1}}
\usepackage[normalem]{ulem}

\begin{document}

\title{Self-similar, spatially localized structures in turbulent pipe flow from a data-driven wavelet decomposition}

\author{Alex Guo\aff{1}, Daniel Floryan\aff{2}\corresp{\email{dfloryan@uh.edu}}, \and Michael D. Graham\aff{1}\corresp{\email{mdgraham@wisc.edu}}}
\affiliation{\aff{1}Department of Chemical and Biological Engineering, University of Wisconsin--Madison, Madison, WI 53706, USA \aff{2}Department of Mechanical Engineering, University of Houston, Houston, TX 77204, USA }



\date{\today}

\maketitle

\begin{abstract}
This study aims to extract and characterize structures in fully developed pipe flow at a friction Reynolds number of $\text{Re}_\tau = 12\,400$. To do so, we employ data-driven wavelet decomposition (DDWD) [D.~Floryan and M.~D.~Graham, PNAS 118, e2021299118 (2021)], a method that combines features of proper orthogonal decomposition and wavelet analysis in order to extract energetic and spatially localized structures from data. We apply DDWD to streamwise velocity signals measured separately via a thermal anemometer at 40 wall-normal positions. The resulting localized velocity structures, which we interpret as being reflective of underlying eddies, are self-similar across streamwise extents of 40 wall units to one pipe radius, and across wall-normal positions from $y^+=350$ to $y/R=1$. Notably, the structures are similar in shape to Meyer wavelets. Projections of the data onto the DDWD wavelet subspaces are found to be self-similar as well, but in Fourier space; the bounds of self-similarity are the same as before, except streamwise self-similarity starts at a larger length scale of $450$ wall units. The evidence of self-similarity provided in this study lends further support to Townsend's attached eddy hypothesis, although we note that the self-similar structures are detected beyond the log layer and extend to large length scales.
\end{abstract}

\section{Introduction}
\label{sec:intro}

While statistical approaches dominated the first half of the 20th century in understanding the dynamics of turbulence, the increasing availability of experimental and direct numerical simulation (DNS) data has enabled structure-based approaches to yield useful insights. For example, in the viscous sublayer of wall-bounded flows, turbulent structures coherent in space and time were discovered and characterized using hydrogen bubble visualization \citep{Kline1967}. These structures consisted of streaks, which are spanwise-alternating regions of high-speed and low-speed fluid, and quasi-streamwise vortices. From this characterization of the sublayer structures, the self-sustaining process (SSP) was developed and is now a widely accepted model for how these structures sustain themselves \citep{Hamilton1995, waleffe1997self, jimenez1999autonomous}. It is evident that the detailed characterization of turbulent structures is a necessary first step toward a mechanistic understanding and modeling of wall-bounded turbulence. Structures in wall-bounded turbulence are reviewed by \cite{robinson1991coherent}, \cite{adrian2007hairpin}, \cite{jimenez2012cascades}, and \cite{McKeon2017}.  

With an eye toward a structural model of wall-bounded turbulence, Townsend proposed the attached eddy hypothesis (AEH) for asymptotically high-Reynolds number flows \citep{Townsend1976}. The AEH supposes that the main, energy-containing motion is made up of contributions from attached eddies. Attached eddies are persistent flow patterns in the log layer whose size scales linearly with their distance from the wall; they span the entire range of scales in the log layer and---critically---are presumed to be geometrically self-similar. The AEH makes specific predictions for how the normal components of the Reynolds stress vary with distance from the wall, which are borne out by experimental \citep{Hultmark2012, hutchins2012towards, winkel2012turbulence, kulandaivelu2012evolution, marusic2013logarithmic} and numerical data \citep{lee2015direct}. Townsend did not commit himself to a particular form for the attached eddies \revise{in his statistical AEH model}. \revise{Inspired by \cite{Kline1967}, Townsend considered a physical, double-cone model; unfortunately, it was inconsistent with the AEH, and Townsend's modeling efforts ended there.} 

Inspired by the flow visualization results of \cite{head1981new}, \cite{Perry1982} specified a shape for the attached eddies, leading to their attached eddy model (AEM). In addition to making the same predictions as the AEH for the normal components of the Reynolds stress, their AEM also yields the log law for the mean flow and predictions for streamwise spectra. They found that all eddy shapes yield similar first- and second-order statistics, but predictions beyond basic statistics are dependent on the shape of the eddy. A key point is that attached eddies, however they look, are spatially localized. More details on the AEM and subsequent refinements are reviewed by \cite{Marusic2019}. 

Partly spurred by the success of the AEH, an increasing abundance of investigations have explored whether self-similar coherent motions are present in wall-bounded turbulence. \cite{Tomkins2003} found that the spanwise length scales of conditionally averaged low-speed structures grew linearly with distance from the wall, as attached eddies are supposed to. \cite{del2006self} found that, on average, the spanwise and streamwise length scales of vortical structures grow roughly linearly with distance from the wall. \cite{lozano2014time} concluded the same for sweep and ejection structures, and \cite{hwang2018wall} concluded the same for structures of positive or negative velocity fluctuations. These studies show that the length scales of various structures grow, on average, with distance from the wall. 

Evocative results were published by \cite{dennis2011experimental}. By conditionally averaging their three-dimensional particle image velocimetry measurements based on spanwise swirl strength at a chosen distance from the wall, they extracted a structure with the shape of a horseshoe vortex. Repeating the analysis at several distances from the wall, they found a similar shape at each distance whose size grew with distance from the wall. The authors suggested that the shape would make a sensible choice of a `representative eddy.' Although the analysis was qualitative and performed in the outer layer, it provided exciting hints of the presence of self-similar coherent structures. 

Recalling that Townsend's eddies constitute the main \emph{energy}-containing motions, \cite{Hellstrom2016} used proper orthogonal decomposition (POD) to explore the existence of self-similar eddies in turbulent pipe flow. These results were also highly evocative as the leading (most energetic) POD modes collapsed on top of each other when scaled by their size. The self-similar POD modes spanned a decade in wall-normal length scales, and the largest structures extended into the outer region. 

Although POD provides an approach to extract energetic structures from data, it has the drawback that in homogeneous directions, the POD modes are constrained to be Fourier modes, precluding direct identification of structure or localization \citep{Holmes2012}. Lumley has espoused the view that such unlocalized structures should not be considered eddies \citep{lumley1970stochastic, tennekes1972first, lumley1981coherent}. A well-known formalism incorporating localization is that of wavelets. A wavelet basis consists of elements localized in both space and scale. Traditionally, the basis elements are formed by translations and dilations of a pre-specified vector called the mother wavelet; this construction makes wavelets perfectly self-similar across scales. Due to their localization and the space-scale unfolding they produce, wavelets have found much use in turbulence \citep{argoul1989wavelet, everson1990wavelet, meneveau1991dual, Meneveau1991, yamada1991identification, farge1992wavelet, katul1994intermittency, katul1998identification, farge2001coherent, farge2003coherent, okamoto2007coherent, ruppert2009wavelet}. More information about wavelets is available in the brief review article by \cite{strang1989wavelets} and in the books by \cite{daubechies1992ten}, \cite{meyer1992wavelets}, \cite{mallat1999wavelet}, and \cite{frazier2006introduction}. 

In applications, a drawback of wavelets is that there are many possible choices for the mother wavelet. Naturally, one may wonder which wavelet is best suited for a particular application. \cite{katul1996partitioning} considered this question in the context of atmospheric turbulence. Motivated by the AEH, they sought to decompose flow variables into contributions from attached (energy-containing) eddies and detached (negligible-energy) eddies. \citeauthor{katul1996partitioning} used an entropy-minimizing scheme developed by \cite{coifman1992entropy} to select the optimal wavelet from a library of known wavelets. We emphasize that the shapes of the wavelets are still pre-specified and that they are perfectly self-similar across different scales. These two features---common in any traditional wavelet analysis---are undesirable if one's goal is to extract structure from data and to determine the degree of self-similarity of the data. 

Motivated by the preceding discussion, \cite{Floryan2021} developed a technique called data-driven wavelet decomposition (DDWD). DDWD integrates the data- and energy-driven nature of POD with the space and scale localization properties of wavelets without imposing self-similarity across scales. \cite{Floryan2021} applied DDWD to homogeneous isotropic turbulence and found that the structures in the data (represented by the data-driven wavelets) were self-similar for scales corresponding to the inertial subrange, reminiscent of the Richardson cascade. Our goal in this work is to leverage DDWD as a data-driven basis of energetic, spatially localized, multiscale vectors for characterizing the shape and self-similarity of structures in wall-bounded turbulence. We extend and apply the technique to experimental measurements in high-Reynolds number pipe flow. As we will show, we identify strongly self-similar, spatially localized, energetic structures across a range of scales---candidate eddies. We describe the experimental setup and measurements in Section~\ref{sec:exp}. We then briefly review DDWD in Section~\ref{sec:ddwd} and describe extensions to accommodate the presently analyzed experimental measurements. Section~\ref{sec:res} describes the structures that we identify, and we end by placing our results in the context of the literature in Section~\ref{sec:lit_compare}.

\section{Experimental setup}
\label{sec:exp}

The analysis in this work is based on data acquired in the Princeton Superpipe facility, a facility capable of creating fully developed pipe flow at high Reynolds numbers. The data were acquired at a Reynolds number of $Re_D = U_b D / \nu = 608\,000$, with a corresponding friction Reynolds number of $Re_\tau = u_\tau R / \nu = 12\,400$, using a cross-wire thermal anemometry probe. Here, $U_b$ is the bulk velocity, $D = 2R$ is the pipe diameter, $\nu$ is the kinematic viscosity of the fluid, and $u_\tau = \sqrt{\tau_w / \rho}$ is the friction velocity, where $\tau_w$ is the wall shear stress, and $\rho$ is the density of the fluid. The pipe is hydraulically smooth for this operating condition \citep{mckeon2004friction}. Details of the facility can be found in \cite{zagarola1998mean}, and details of the data we use and the measurement probe can be found in \cite{fu2019design}; we recapitulate below. 

\begin{figure}
    \centering
    \includegraphics[width=0.9\textwidth]{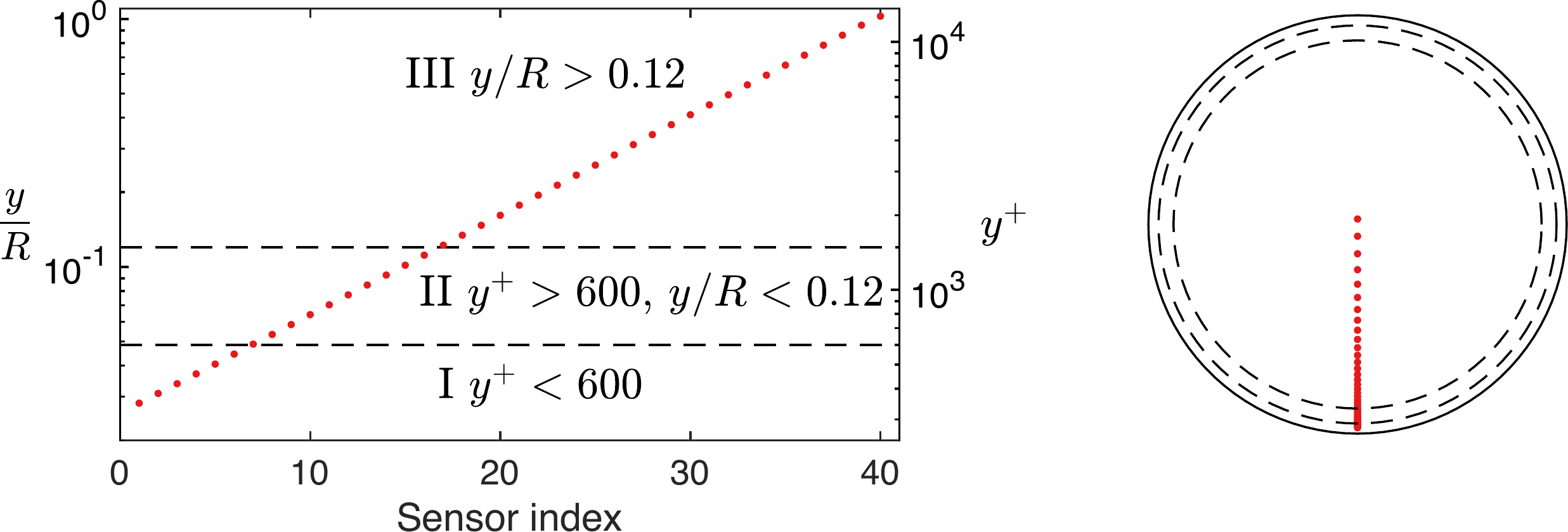}
    \caption{Wall-normal positions of the measurement probe. }
    \label{fig:sensorLoc}
\end{figure}

The data consist of long time series of the streamwise velocity component measured at 40 distances from the pipe wall, with two time series taken at each distance. The measurements at different distances from the wall were not simultaneous. The positions $y$ of the measurement probe are shown in Figure~\ref{fig:sensorLoc}, with a superscript + denoting wall units (non-dimensionalized by $\nu / u_\tau$). The measurements were acquired in three regions of the flow: (I) the power-law region, (II) the log-law region, and (III) the wake region. Although the log-law region is classically taken to start at $y^+ = 30$ \citep{pope2000turbulent}, experiments in the Superpipe suggest that the mean streamwise velocity follows a power law for $50 < y^+ < 600$, with the log law not taking hold until $y^+ = 600$ \citep{mckeon2004further}.

Each time series consists of roughly $1.67 \times 10^7$ samples, sampled at $F_s=300$ kHz, and filtered using an analog 8-pole Butterworth filter at 150 kHz. We convert the temporal signals to spatial signals using Taylor's frozen field hypothesis, in which case distances between consecutive samples range from 2.6 wall units at the measurement position closest to the wall to 4.0 wall units at the measurement position furthest from the wall. The measurement probe had a volume of 42 $\mu$m $\times$ 42 $\mu$m $\times$ 50 $\mu$m, or 8.0 $\times$ 8.0 $\times$ 9.5 wall units$^3$. The length of the wires in the measurement probe was 60 $\mu$m, or 11.4 wall units. 

\begin{figure}
    \centering
    \includegraphics[width=0.5\textwidth]{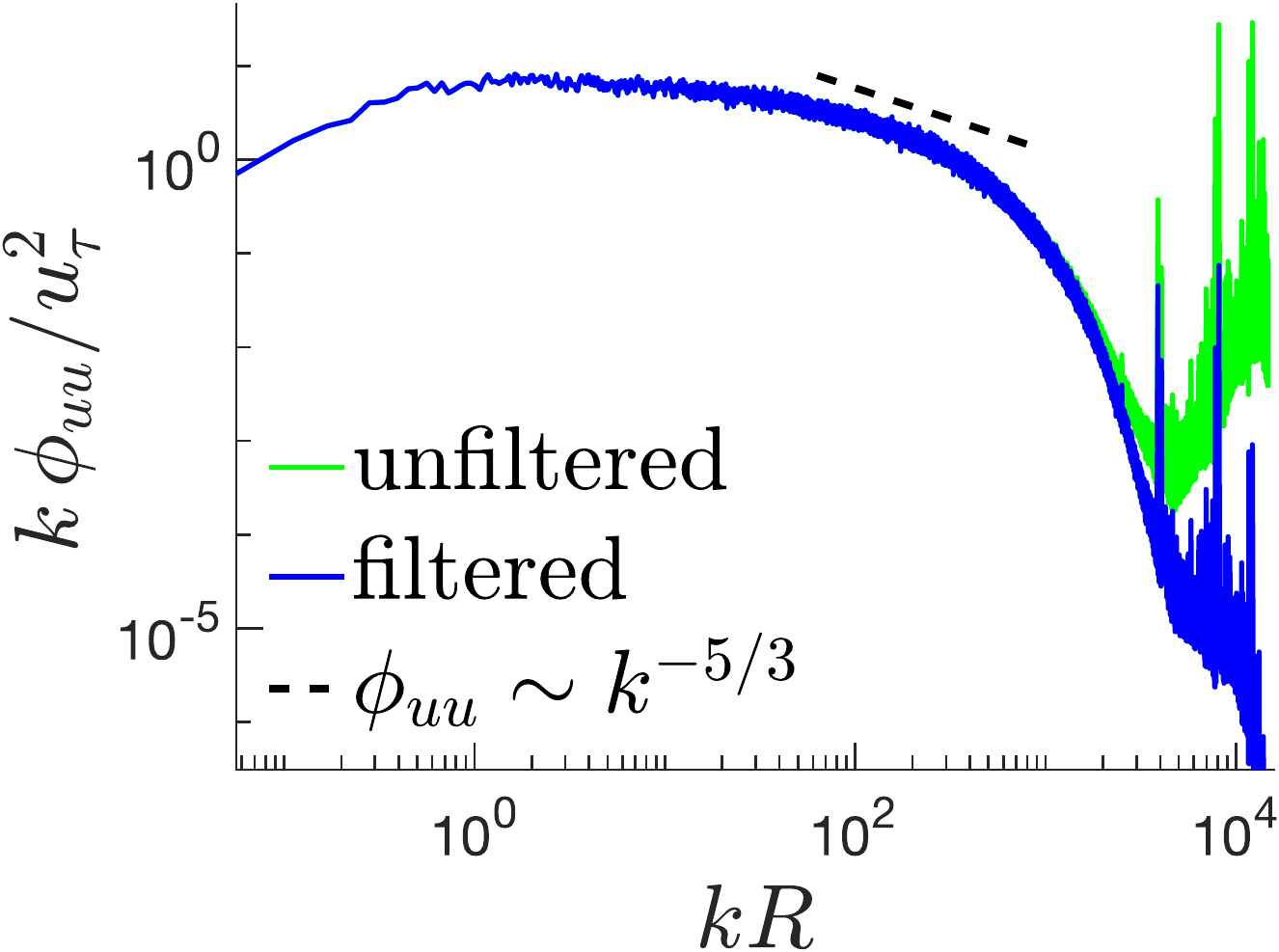}
    \caption{Premultiplied PSD of $u$, measured at $y^+=350$, without and with filtering at 25 kHz.}
    \label{fig:filtered_PSD}
\end{figure}

Since the time series were filtered at the Nyquist frequency of 150 kHz during collection, the energy at the largest wavenumbers $k$ corresponding to the measurement noise remains unattenuated. We filter each time series again with a 2-pole Butterworth filter at 25 kHz. Figure \ref{fig:filtered_PSD} shows the effect of this filter on the premultiplied power spectral density (PSD) of the signal measured closest to the wall at $y^+=350$ (the rest of the time series are affected similarly by the same filter). The PSD, $\phi_{uu}$, was estimated using segments that contained $2^{19}$ elements, overlapped by 50\%, and windowed with the Hamming window.

\begin{figure}
    \centering
    \includegraphics[width=0.6\textwidth]{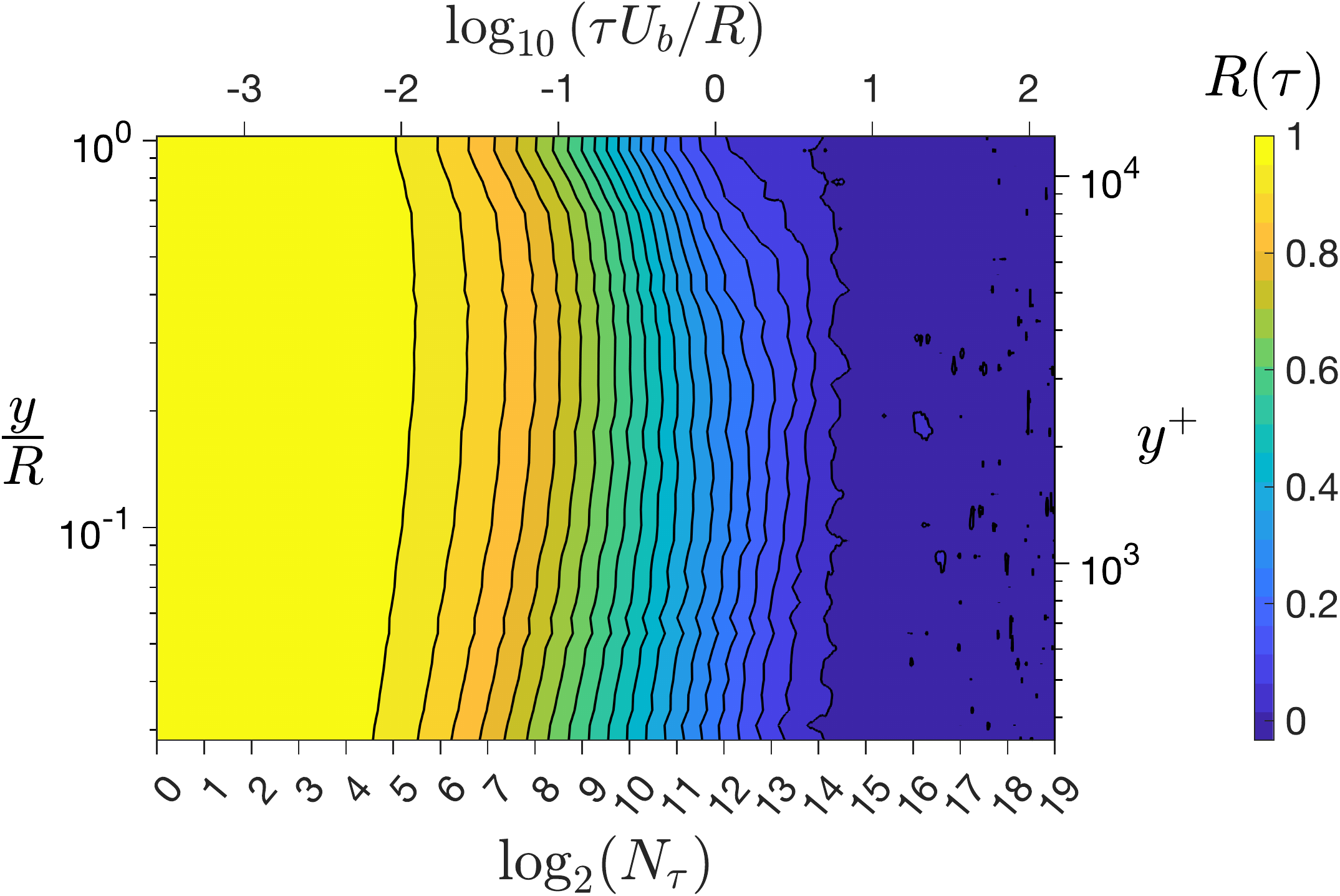}
    \caption{Heat map of the time correlation function at all wall-normal positions. $N_\tau = \tau F_s$ is the number of points needed to capture a time lag of $\tau$ [s], and $\tau U_b/R$ is the dimensionless eddy turnover time.}
    \label{fig:autocorr_heatmap}
\end{figure}

Before describing the method we will use to analyze the structure of the Superpipe data, we estimate the largest coherent structure---or eddy---we can expect to find. The largest eddies in a flow are usually referred to as the very-large-scale motions (VLSMs), and their sizes can be obtained from the PSD of the streamwise velocity. Specifically, the initial peak in the premultiplied PSD is taken to correspond to the VLSM \citep{kim1999very}. Selecting this peak can be difficult, however, since producing a good estimate of the PSD involves a tradeoff between using longer data segments (yielding greater resolution in $k$) and more segments (yielding less variance in the PSD estimate). Even though our time series are rather long, the PSD in Figure \ref{fig:filtered_PSD} is the best we can produce without relying on using a moving-average filter in $k$-space. Estimating the location of the initial peak in the premultiplied PSD \revise{can} be fraught with uncertainty. To avoid the tradeoff involved in estimating PSDs, we opted to use the time correlation function to estimate the largest expected coherent structure. The heatmap of the time correlation function at all 40 wall-normal positions is shown in Figure \ref{fig:autocorr_heatmap}, and it suggests all 40 time series are correlated up until 10 eddy turnover times. Since $U_b=5.4$ m/s and the mean velocity $\overline U$ only ranges from 4.2 to 6.4 m/s over the range of wall-normal positions considered here, one can effectively say the structures at all wall-normal positions are correlated in the streamwise direction up until $\mathcal O(1 R)$ to $\mathcal O(10 R)$. VLSMs have previously been estimated to be $\mathcal{O}(10 R)$ in length \citep{kim1999very}. 

\section{Data-driven wavelet decomposition}
\label{sec:ddwd}

Data-driven wavelet decomposition was introduced by \cite{Floryan2021}, and has its roots in POD and wavelet analysis. First, we provide a conceptual overview of DDWD and its connections to POD and wavelet analysis---for full details, we refer the reader to \cite{Floryan2021} and to Appendix \ref{sec:appendix}. We then describe the pre-processing steps and modifications needed to analyze the long time series described in Section~\ref{sec:exp}.

\subsection{Overview and connections to POD and wavelet analysis}
\label{sec:overview}

We consider periodic vectors $z\in \mathbb{R}^N$, with $N$ even, and a dataset $\{z_i \in \mathbb{R}^N\}_{i=1}^{M}$. Wavelets, POD, and DDWD construct different orthonormal bases for $\mathbb{R}^N$. We interpret the basis elements as structures. Wavelet basis elements are localized structures that are independent of the data being analyzed. POD basis elements, on the other hand, tend to be global structures, and they derive from the data under consideration. Specifically, POD basis elements are constructed via an energy maximization problem and are thus interpreted as energetic structures present in the data. DDWD combines the main features of wavelets and POD, producing basis elements representing energetic, localized structures present in the data. 

\begin{figure}
    \centering
    \includegraphics[width=0.6\textwidth]{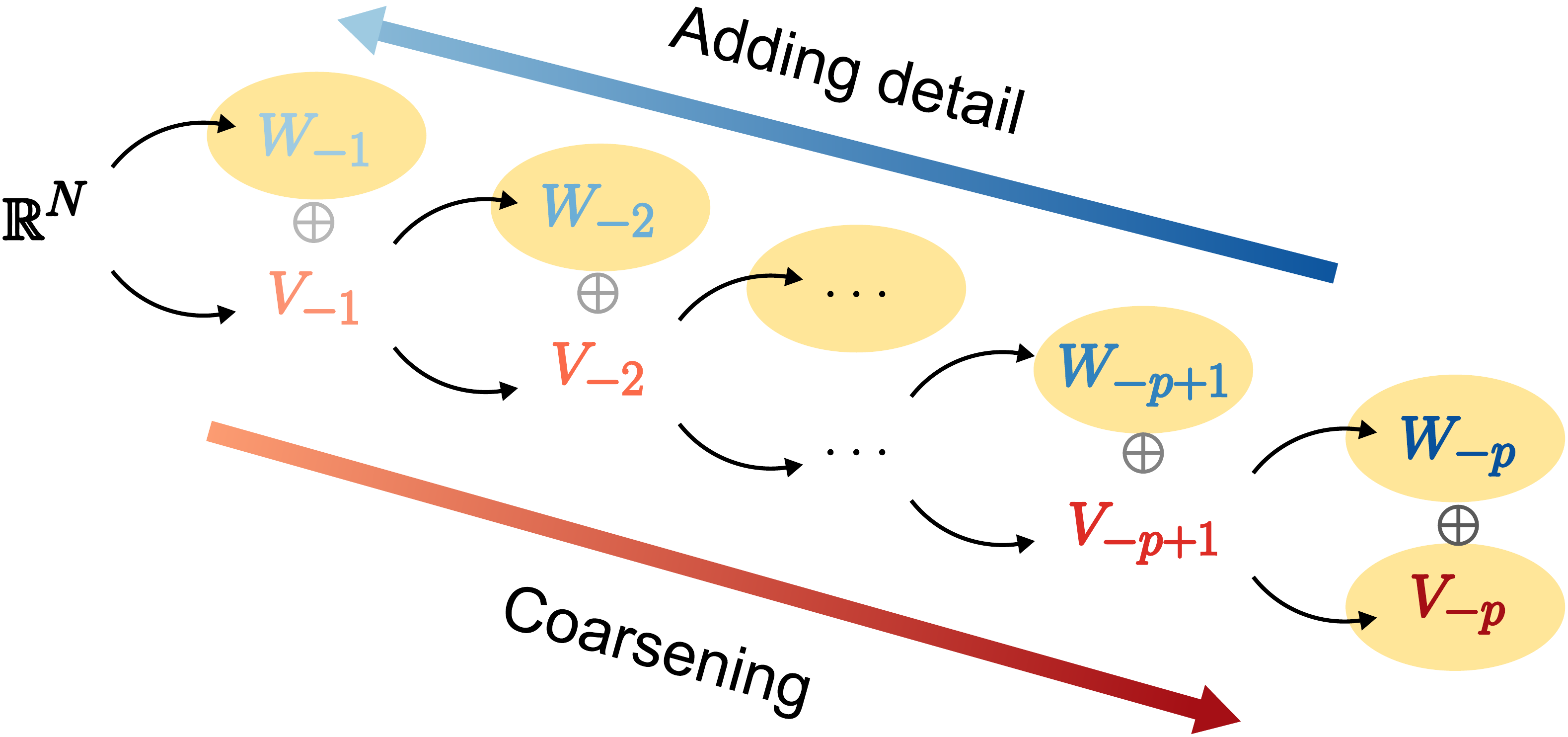}
    \caption{Subspaces formed by wavelets periodic in $\mathbb R^N$. The full space $\mathbb R^N$ is composed of the highlighted subspaces.}
    \label{fig:subspaces}
\end{figure}

\begin{figure}
    \centering
    \includegraphics[width=0.9\textwidth]{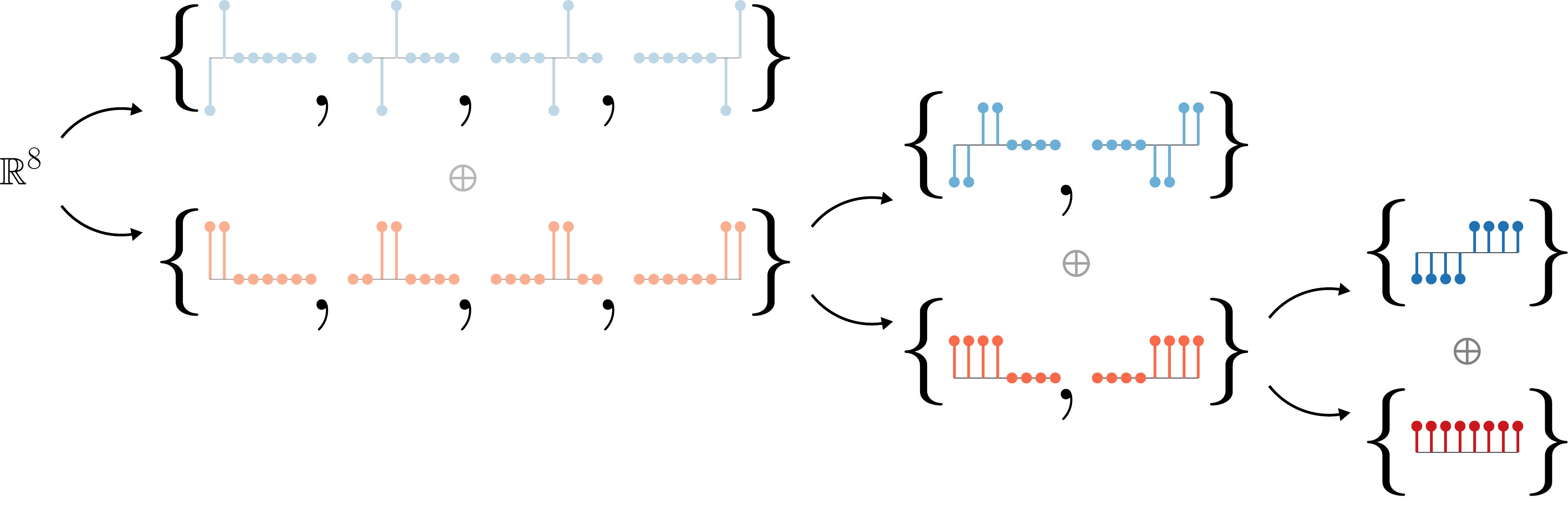}
    \caption{Haar wavelets on $\mathbb{R}^8$, following the style of Figure~\ref{fig:subspaces}. The Haar basis for each subspace is shown. }
    \label{fig:haar}
\end{figure}

For our purposes, the most useful way to think about wavelets is by the associated subspaces. A wavelet basis splits $\mathbb{R}^N$ into two orthogonal subspaces of dimension $N/2$: the approximation subspace $V_{-1}$, and the detail subspace $W_{-1}$. Subspace $V_{-1}$ is spanned by $N/2$ mutually orthonormal even translates of the father wavelet, $\phi_{-1}$, and subspace $W_{-1}$ is spanned by $N/2$ mutually orthonormal even translates of the mother wavelet, $\psi_{-1}$. \revise{(By ``even'', we mean the translates are shifted by an even number of mesh points.)} The mother wavelet can be constructed from the father wavelet, and vice versa. Traditionally, projecting a vector in $\mathbb{R}^N$ onto the approximation subspace $V_{-1}$ yields a low-pass filtered version of the vector, and projecting the vector onto the detail subspace $W_{-1}$ yields a high-pass filtered version of the vector (hence the nomenclature of ``approximation'' and ``detail'' subspaces). Since $V_{-1}$ and $W_{-1}$ are orthogonal complements, the vector is equal to the superposition of these two projections. The splitting of subspaces is performed recursively on the approximation subspace, ultimately yielding a decomposition $\mathbb{R}^N = V_{-p} \oplus W_{-p} \oplus \ldots \oplus W_{-1}$ for $N = 2^p$, illustrated in Figure~\ref{fig:subspaces}. One may think of this decomposition as the direct sum of a very coarse approximation subspace ($V_{-p}$) and progressively finer detail subspaces ($W_{-p}$ to $W_{-1}$). At each stage $l$, $V_{-l}$ and $W_{-l}$ are respectively spanned by $N/2^l$ mutually orthonormal translates of $\phi_{-l}$ and $\psi_{-l}$. In a traditional wavelet basis, one specifies the mother wavelet, $\psi_{-1}$, and all other basis elements derive from the mother wavelet. An example---the Haar wavelet basis---is shown in Figure~\ref{fig:haar}. Note two salient features of wavelet bases: spatial localization of the basis elements, and self-similarity across stages (the wavelets at coarser stages are essentially simple dilations of the mother wavelet).

POD can be considered in a similar light as wavelet analysis. Letting $\{u_j\}_{j=1}^N$ denote the ordered set of POD basis elements, we have the following decomposition: ${\mathbb{R}^N = \text{span}(u_1) \oplus \ldots \oplus \text{span}(u_N)}$. That is, POD decomposes $\mathbb{R}^N$ into a direct sum of one-dimensional subspaces. Whereas wavelet subspaces are ordered by scale, POD subspaces are ordered by energy content; whereas wavelet subspaces are determined by the choice of the mother wavelet, POD subspaces are determined from data. 

DDWD combines POD and wavelet analysis. Namely, it maintains the subspace structure of a wavelet basis (shown in Figure~\ref{fig:subspaces}, where each subspace is spanned by mutually orthonormal translates of a single wavelet), but the wavelet at each stage is determined from data by an energy maximization principle. Letting $E(\cdot)$ denote the energy of a dataset contained in a subspace, orthogonality of the subspaces implies that $E(V_{-l}) = E(V_{-l-1}) + E(W_{-l-1})$. That is, whenever a subspace is split, the energy contained in it is also split. Motivated by the observation that large-scale (or coarse) structures tend to be the most energetic ones in fluid flows, each time we split a subspace, we design the subsequent subspaces so that the approximation subspace contains as much of the dataset's energy as possible. An analogous approach is to progressively minimize the energy in the detail subspaces. The design of the subspaces can be framed as an optimization problem. Consider an ensemble of \revise{$M$} data vectors forming the dataset $Z = [z_1 \ z_2 \ \hdots \ z_M] \in \mathbb R^{N \times M}$. The mathematical formulation for finding the first-stage DDWD wavelet is
\begin{equation}
    \label{eq:opt}
    \begin{array}{cc}
        \min \limits_{\psi_{-1}} & \frac{1}{\| Z \|_F^2} \underbrace{ \sum \limits_{m=0}^{N/2-1} \| Z^T R^{2m}(\psi_{-1}) \|^2 }_{M E_1} + \lambda^2 \text{Var}(\psi_{-1}) \\\\ 
        \text{s.t.} & \psi_{-1}^T R^{2m}(\psi_{-1}) = \delta_{m 0} , \quad m = 0, \hdots, N/2-1 ,
    \end{array}
\end{equation}
where $R^q$ is the linear operator that circularly shifts its input by $q$ (e.g., for $\psi_{-1} = [a,b,c,d]^T$, $R^1(\psi_{-1}) = [d,a,b,c]^T$). Above, $E_1$ denotes the mean energy of the data vectors projected onto $W_{-1}$\revise{, and $F$ denotes the Frobenius norm}. The first term in~\eqref{eq:opt} is the fraction of the dataset's energy in the finest detail subspace (which typically contains the smallest scale). The circular variance of the wavelet, which is bounded between 0 and 1, is penalized in the second term to encourage the wavelet to be spatially localized. The results are robust to the strength of the penalty parameter $\lambda^2$, with the penalty term hardly affecting the wavelets' ability to capture energy \citep{Floryan2021}. The reason is that the penalty term manipulates the phases of each Fourier component of a wavelet to achieve spatial localization while essentially leaving the magnitudes untouched. Moreover, we find that the shape of the wavelet found by DDWD is generally robust to the choice of $\lambda^2$ across several orders of magnitude (see Appendix~\ref{sec:app_synth}). For the results in this work, we have used $\lambda^2 = 10^{-4}$. 

The wavelets at later stages are found recursively. To find $\psi_{-2}$, we project the dataset onto $V_{-1}$, replace $N$ by $N/2$ and $R^{2m}$ by $R^{4 m}$ in~\eqref{eq:opt}, and decrease $\lambda^2$ by a factor of 4 \citep{Floryan2021}. In the end, DDWD produces an energetic hierarchy of spatially localized orthonormal structures.

We emphasize two differences between DDWD and traditional wavelet analysis. The first is that the concept of energy is built into DDWD. Whereas the stage of the decomposition induced by a traditional wavelet basis is synonymous with the scale of the wavelets, this is not necessarily the case for a basis produced by DDWD since energy is also taken into account. Scale and energy are often related in fluid flows, however, in which case the energetic hierarchy of structures found by DDWD will also be a hierarchy of scales. 

The second difference is that self-similarity across stages is not built into DDWD. It bears repeating that only energy is taken into consideration when finding structures at each stage of the decomposition, and a separate optimization problem is solved at each stage. For data that are not self-similar, neither are the basis elements produced by DDWD. The converse is also true: self-similarity in the data manifests itself in self-similar basis elements. In \cite{Floryan2021}, DDWD applied to data from homogeneous isotropic turbulence produced self-similar basis elements in the range of scales corresponding to the inertial subrange, while when it was applied to white noise and data generated by the Kuramoto–Sivashinsky equation, the basis elements were not self-similar. Determining the degree of self-similarity of structures in the Superpipe data will feature prominently later in this work.

\subsection{Pre-processing and modifications for long time series}
\label{sec:mod}

Like POD, DDWD uses an ensemble of vectors (the data). Here, however, we have two long time series of data (for each wall-normal position, which will be analyzed separately). We form an ensemble by breaking the two time series into many shorter segments, as is done when estimating the PSD. Doing so introduces a number of issues that must be addressed: how long should the shorter segments be; how can border effects be minimized; and how can we efficiently capture very small and very large structures?

\subsubsection{Data segmentation}
\label{sec:seg}

We first address how to segment a long time series into an ensemble of shorter segments of length $N=2^p$. The segments need to be long enough to capture the length of the largest coherent structure---or eddy---in the flow. Figure \ref{fig:autocorr_heatmap} suggests we need $p \ge 15$ to capture the largest correlated structure; however, the last 2--3 stages of any wavelet basis contain wavelets that are not spatially localized. We set $p=18$ to ensure the largest spatially localized wavelet represents the largest correlated/coherent structure.

\subsubsection{Border effects}
\label{sec:border}

Following the classical discrete wavelet transform, DDWD implicitly assumes that signals are periodic.  Of course, this is not true for our segmented data. If a segment of data is treated as periodic, there will generally be a discontinuity when the end of the segment is wrapped back around to the beginning. A wavelet decomposition of a signal with a discontinuity generally has high energy in the small-scale wavelets localized around the discontinuity. This border effect could greatly distort the data-driven wavelets found by DDWD. 

To circumvent distortions due to border effects, we turn to PSD estimators. PSD estimators based on the discrete Fourier transform also implicitly assume periodic signals, so they are also subject to border effects. To minimize border effects, estimators window the data vectors before taking discrete Fourier transforms of them \citep{press2007numerical}. Windowing data consists of multiplying the data by a function $w$ that rises from zero to a peak of one and then falls back to zero. We do the same, windowing our data vectors with a 50\% Tukey window before performing DDWD on them. \cite{Meneveau1991} similarly windowed his non-periodic data before performing a wavelet analysis. We have tested this procedure on windowed and unwindowed periodic data, finding that windowing does not affect the shapes of the data-driven wavelets. The type of window used also does not have a significant impact. We double the number of data vectors by overlapping the segments by 50\%; this is a common practice to decrease variance when estimating spectra \citep{press2007numerical} and doing so allows almost every part of the signal to appear in our dataset unattenuated by the 50\% Tukey window.

Since windowing data attenuates the signal, PSD estimators compensate for this attenuation in order to calculate accurate estimates of the energy contained within each Fourier mode. For a window $w \in \mathbb R^N$, the attenuation caused by the windowing is counteracted by dividing the energy in each Fourier mode by $w^T w/N$, which is bounded between zero and one \citep{press2007numerical}. Each DDWD subspace contains basis elements that (as a whole) span the entire spatial domain, just like individual Fourier modes. Consequently, we can account for the attenuation from the windowing by dividing the energy in $Z$ and each DDWD subspace by $w^T w/N$. We account for the segment overlapping, which doubles our number of data vectors, by further dividing the energy by 2. The additional factor arises for the following reason. For the long time series $u$ (before segmenting it), its energy is $u^Tu$. After we segment it with 50\% overlap, we store the resulting segments in a data matrix $Z$. Each entry of $u$ appears twice in $Z$ due to the 50\% overlap (except entries near the beginning and end of $u$). Then the energy in $Z$ is $\Vert Z \Vert_F^2 \approx 2 u^T u$.
Note that multiplying the data by a constant does not affect the optimization problem in~\eqref{eq:opt} because of how the first term in the objective function is normalized. Hereafter, $l$ will denote the stage and $E_l$ will denote the stage-$l$ energy with the factor of $2 w^T w/N$ accounted for.

\subsubsection{Capturing disparate scales}
\label{sec:scales}

An important issue is how to efficiently capture very small and very large scales. The very smallest scales in the data will be of a size on the order of one element (2.6 to 4.0 viscous lengths, depending on the wall-normal position) in our data segments, and the very largest scales will be of a size on the order of $2^{18}$ elements ($55R$ to $85R$). The constrained optimization problem in~\eqref{eq:opt} is converted to an unconstrained optimization problem (see \citep{Floryan2021}) and solved using the BFGS algorithm. The optimization problem scales poorly since evaluating the converted objective function once has computational complexity $\mathcal{O}(N^2 \log_2 N + N^2 M)$ and the number of iterations it takes the BFGS algorithm to converge grows with $N$ (although the rate of growth is unclear). This makes the problem intractable for data vectors of length $2^{18}$. 

We solve this problem by exploiting the iterative nature of a wavelet decomposition and the spatial localization of data-driven wavelets. In the first stage of DDWD, we find the lowest-energy (typically finest-scale) structures. Since these first-stage structures tend to be highly localized, we do not need long data vectors to capture them. Thus, we can perform DDWD on an ensemble of short vectors of length $N_s \ll N$ to capture the smallest-scale structures. 

The result of the first stage of DDWD will be wavelets of length $N_s$, which form part of a basis for $\mathbb{R}^{N_s}$. Ultimately, we need to work our way up to full-length vectors of length $N$ since we seek a basis for $\mathbb{R}^N$. Taking advantage of the spatial localization of the wavelets, we simply pad them with zeros until their length is $N$. With the full-length first-stage wavelets in hand, we use them to coarsen the full-length data vectors by projecting the data onto the approximation subspace $V_{-1}$ (see Figure~\ref{fig:subspaces}). The projected data have length $N/2$ and are simply coarsened versions of the original data. We proceed iteratively, performing one stage of DDWD at a time on shortened segments of the coarsened data. The technical details are in Appendix \ref{sec:app_18s}.

Note that once we have our data-driven wavelets, performing a wavelet decomposition with them scales well with the length of the data vectors. It is only the algorithm used to find the data-driven wavelets that scales poorly. 


\subsubsection{Enforcing zero mean and validating against synthetic data}
\label{sec:enforce}
In traditional wavelet bases, $\phi_{-18}$ from stage $p=18$ is always a vector of ones and represents the mean component of all elements in $\mathbb R^N$. Due to orthogonality, all wavelets $\psi_{-l}$ and their translates must have zero mean. When using DDWD to generate a wavelet basis for a given dataset, one can introduce a hard constraint to enforce zero mean on $\psi_{-l}$ \citep{Floryan2021}; however, this hard constraint is generally not needed since the energetically minimal $\psi_{-l}$ that DDWD finds also tends to have nearly zero mean. Nonetheless, we choose to enforce zero mean on $\psi_{-l}$ in this work to mitigate a numerical issue that causes the wavelets in stages 13 and up to have artificially high wavenumber content. This numerical issue was found by validating DDWD against synthetic signals composed of known wavelets (see Appendix~\ref{sec:app_synth} for details). Synthetic data were also used to demonstrate DDWD's ability to uncover many types of wavelets, its robustness to the variance penalty parameter $\lambda^2$ across multiple orders of magnitude, the necessity of windowing, and the validity of the modification to DDWD described in Section \ref{sec:scales}.

\section{Results}
\label{sec:res}

We apply DDWD to the Superpipe data described in Section~\ref{sec:exp}. Starting with the measurements made nearest to the wall ($y^+ = 350$), in Sections~\ref{sec:shape} and~\ref{sec:self_sim} we respectively examine the streamwise shape and degree of self-similarity of the localized structures we compute. Extending the analysis to all wall-normal positions, in Section~\ref{sec:sim_y} we discuss the computed structures' streamwise self-similarity and the similarity across different wall-normal positions. In Section~\ref{sec:wave_proj} we examine projections of the data onto the computed wavelet subspaces. Lastly, in Section \ref{sec:energy} we derive how spectral power law scalings translate to wavelet space and search for the presence of such scaling laws.

\subsection{Shape of localized structures at $y^+ = 350$}
\label{sec:shape}

\begin{figure}
    \centering
    \includegraphics[width=1\textwidth]{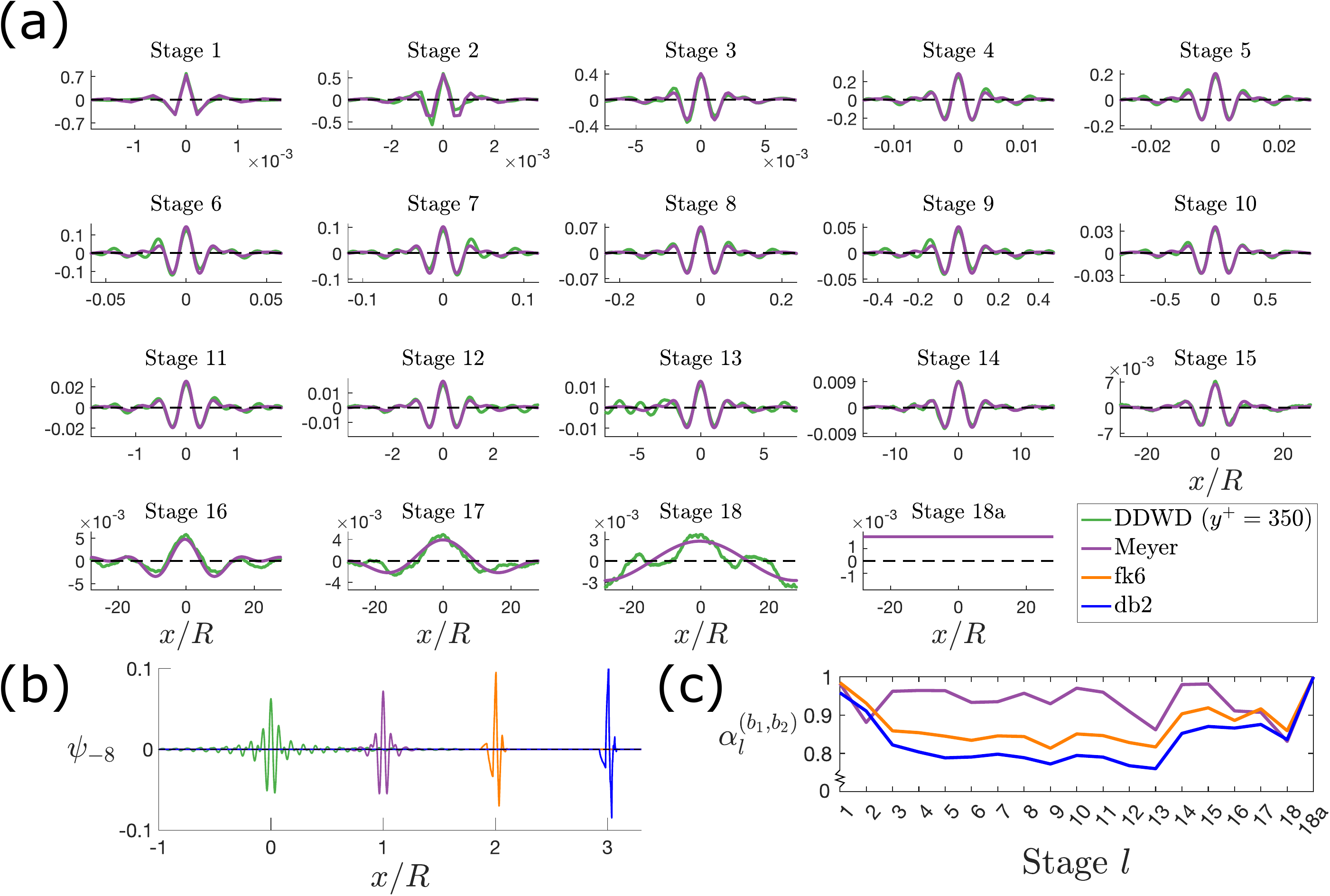}
    \caption{Comparing the 18-stage DDWD basis at $y^+=350$ to three known wavelet bases. (a) Superimposed wavelets (only DDWD and Meyer shown for clarity). (b) Stage-8 wavelets. (c) Similarity between DDWD basis and known wavelet bases.}
    \label{fig:wave_compare}
\end{figure}

Figure~\ref{fig:wave_compare}a shows the wavelets contained in the 18-stage DDWD basis. For comparison, we have also plotted Meyer wavelets. Each plot corresponds to a different stage of the wavelet decomposition, showing the corresponding wavelet. The plot corresponding to ``Stage 18a'' shows the basis element for the coarsest approximation subspace, $V_{-18}$. For the first 15 stages, the range of the horizontal axis is progressively dilated by a factor of two. For ease of comparison, we have shifted the wavelets and, in some cases, reflected them about their horizontal axes \revise{(equivalent to multiplying the wavelet coefficients by $-1$)}. \revise{Although a stage-$l$ wavelet only has translational invariance with respect to shifts by $2^l$ (i.e., the spacing between its translates), shifting it by any integer is justifiable here since the translational invariance of the data causes the energy contained in the subspace $W_{-l}$ to remain the same no matter the shift value.} We observe that the scale and stage of the DWWD wavelets are synonymous. Note that for stages 15 to 18, the wavelets (from all bases) are not spatially localized or self-similar because their scale approaches the length of the data segments. 


\noindent

Figure \ref{fig:wave_compare}b shows the shapes of the stage-8 DDWD wavelet and the Meyer, Fej\'er-Korovkin 6 (fk6), and Daubechies 2 (db2) wavelets\revise{; we chose these wavelets for comparison since they have varying length scales (which we will quantify later) even within the same stage}. Figure \ref{fig:wave_compare}c quantifies the similarity of these three known wavelets to the DDWD wavelet with the inner product
\begin{equation}
    \label{eq:alpha_fixedl}
    \alpha_l^{(b_1,b_2)} = \left\langle {\psi_{-l}^{(b_1)},\psi_{-l}^{(b_2)}} \right\rangle,
\end{equation}
where $\psi_{-l}^{(b_1)}$ is the stage-$l$ wavelet belonging to basis $b_1$ and similarly for $\psi_{-l}^{(b_2)}$ (in Figure \ref{fig:wave_compare}c, $b_1$ is always the DDWD basis while $b_2$ changes between the three known wavelet bases). The absolute value of the inner product is bounded between 0 and 1. The DDWD wavelets obtained from the Superpipe data are quantitatively most similar to Meyer wavelets. This similarity was also noted when wavelets were computed from data from homogeneous isotropic turbulence \citep{Floryan2021}. \revise{Two potential reasons that the most energetic wavelets in these two datasets are similar to Meyer wavelets, as opposed to fk6 or db2, is that Meyer wavelets are smoother and have a larger length scale at any given stage. Velocity fields are smooth, so smooth basis functions should give a better representation. As for the larger length scale, that may be a consequence of the combination of smoothness and the orthogonality condition, or may be connected to the fact that there is always a nonlocal aspect to incompressible flows because of the global pressure constraint.}

Note that at a given stage, wavelets across different bases can appear very different in shape and width but still have an inner product close to 1 due to having similar Fourier content. In what follows, we consider shapes to be similar when $\alpha_l^{(b_1,b_2)} \gtrsim 0.95$.

\subsection{Self-similarity of localized structures at $y^+ = 350$}
\label{sec:self_sim}

\begin{figure}
    \centering
    \includegraphics[width=1\textwidth]{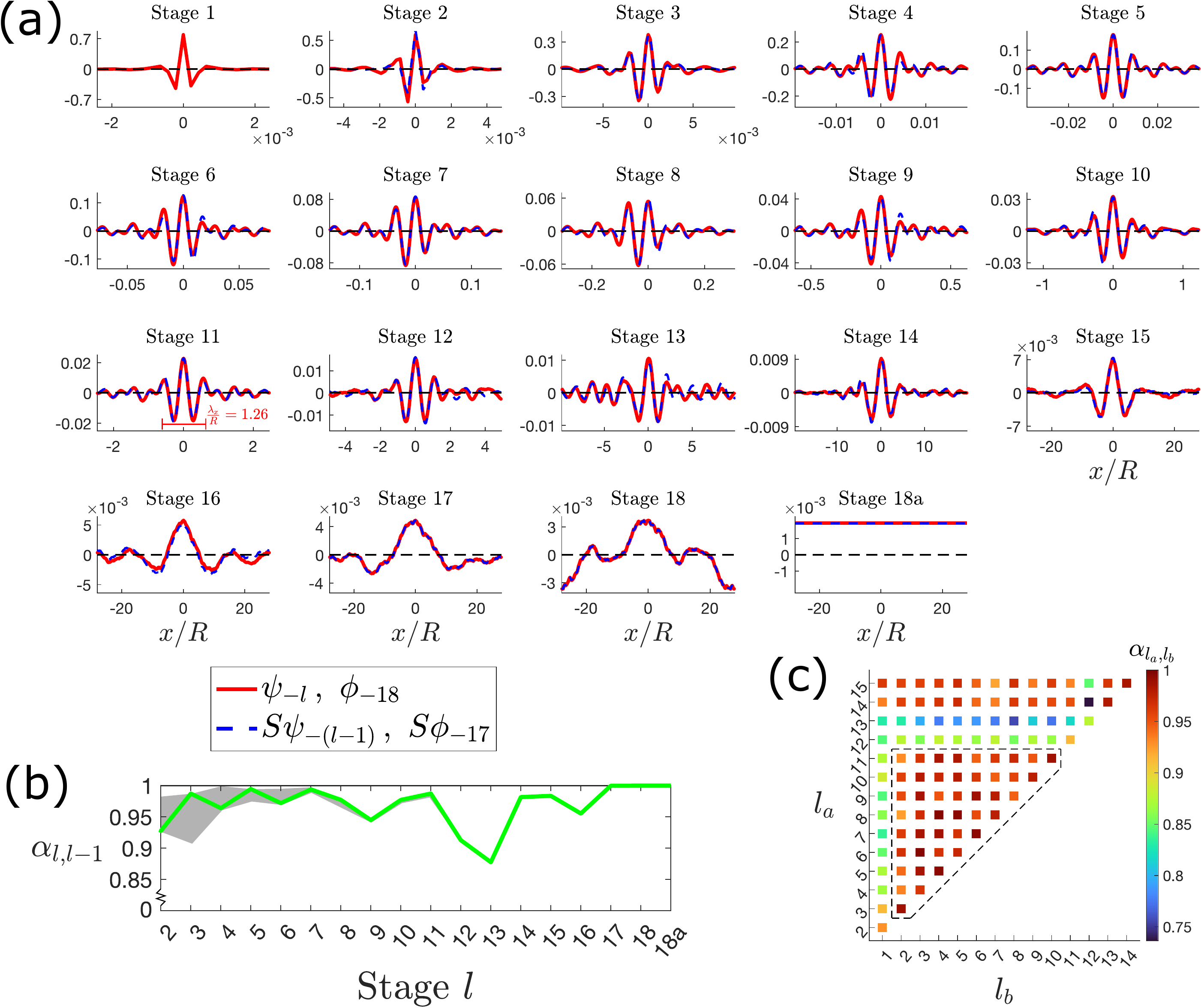}
    \caption{(a) 18-stage DDWD basis at $y^+=350$; solid red curves are $\psi_{-l}$ (or $\phi_{-18}$ for stage $18a$), and dashed blue curves are $S \psi_{-(l-1)}$ (or $S \phi_{-17}$ for stage $18a$), where $S$ essentially dilates its input wavelet by a factor of 2. (b) Self-similarity across adjacent stages (i.e., inner product between the normal and dilated wavelets), where the gray region covers the range of values encountered across 10 DDWD initializations. (c) Self-similarity across stages 1 through 15, where the dashed triangle indicates the region of high self-similarity ($\alpha_{l_a,l_b} \gtrsim 0.95$).}
    \label{fig:self_sim_x}
\end{figure}

We analyze the similarity of DDWD wavelets across stages, which reflects the streamwise self-similarity of the velocity signal. To do so, we must introduce the scaling operator $S$. The scaling operator is used when constructing a traditional wavelet basis, with $\psi_{-l} = S\psi_{-(l-1)}$ for all stages. The scaling operator defines our notion of self-similarity: we say that wavelets across a stage are exactly self-similar when $\psi_{-l} = S\psi_{-(l-1)}$, as is the case for traditional wavelets. This operator essentially dilates its input wavelet by a factor of 2, as is evident by comparing the Meyer wavelets from one stage to the next in Figure~\ref{fig:wave_compare}a (a precise mathematical definition of $S$ is given in \cite{Floryan2021}). In Figure \ref{fig:self_sim_x}a, the stage-$l$ DDWD wavelets $\psi_{-l}$ are plotted with solid red curves and the dilated, previous-stage wavelets $S\psi_{-(l-1)}$ are plotted with dashed blue curves. The wavelets are apparently strongly self-similar across adjacent stages.
\noindent

To quantify the degree of self-similarity, we compute the inner product
\begin{equation}
    \label{eq:alpha_fixedb}
    \alpha_{l_a,l_b} = \left\langle {\psi_{-l_a},S^{l_a-l_b} \psi_{-l_b}} \right\rangle \,, \ l_a \geq l_b,
\end{equation}
analogous to \eqref{eq:alpha_fixedl}.
Figure \ref{fig:self_sim_x}b shows $\alpha_{l_a,l_b}$ for $l_a = l$ and $l_b = l-1$, which is the inner product between the solid and dashed wavelets appearing in each subplot in Figure \ref{fig:self_sim_x}a. A high degree of self-similarity ($\alpha_{l,l-1} \gtrsim 0.95$) across adjacent stages occurs in stages 2 through 11 and then in stages 14 and 15. \revise{The drop in self-similarity between stages 11 and 14 is physical---spatially localized structures are no longer self-similar at these stages. However, since the wavelets in the last 2--3 stages (stages 16 and beyond here) of any basis span the whole spatial domain, we disregard them hereafter. Finally, we note (but do not show here) that the drop in self-similarity at stage 11 is robust to decreasing the total number of stages from $p=18$ to $p=14$.}

Figure \ref{fig:self_sim_x}c shows $\alpha_{l_a,l_b}$ for $l_a > l_b, l_a \in [2,15]$, quantifying the degree of self-similarity across multiple stages. The results in Figure \ref{fig:self_sim_x}b appear along the main diagonal in Figure \ref{fig:self_sim_x}c. One can see that high self-similarity of $\psi_{-l}$ occurs in stages 2 through 11 across all stages, not just adjacent stages. 

Altogether, DDWD reveals that coherent velocity structures (induced by eddies) are highly self-similar in stages 2 through 11. Defining a length scale $\lambda_x$ for a wavelet as the region centered about the origin that contains $90\%$ of its energy, this range of stages corresponds to $\lambda_x^+ = 20$ through $\lambda_x/R = 1.26$. \revise{(Note, our definition for $\lambda_x$ is justified in Appendix~\ref{sec:app_lambdax} and is unrelated to the Fourier wavenumber $k$.)} Not only does DDWD give direct evidence of streamwise self-similarity of localized structures in wall-bounded turbulence, but it also reveals a surprisingly much larger range of self-similarity when compared to many statistical and structural approaches. Section~\ref{sec:lit_compare} includes a more detailed discussion comparing the results obtained using DDWD to those in the literature.


\subsection{Self-similarity of localized structures in wall-normal direction}
\label{sec:sim_y}
\begin{figure}
    \centering
    \includegraphics[width=1\textwidth]{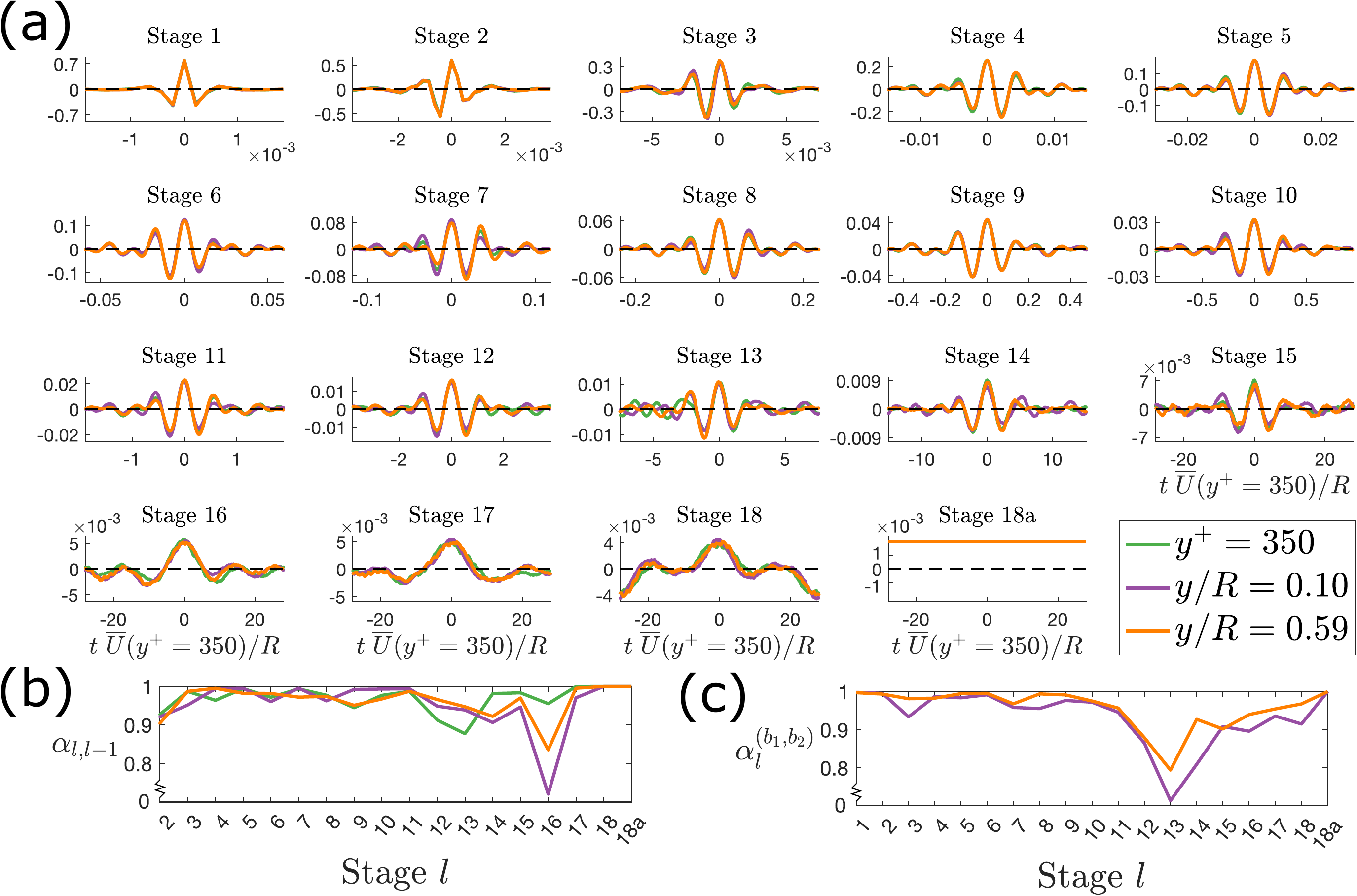}
    \caption{(a) DDWD bases at $y^+=350$, $y/R=0.10$, and $y/R=0.59$. (b) Similarity across adjacent stages for each basis. (c) Similarity across bases, where $b_1$ is for the signal closest to the wall and $b_2$ is for the signal either in region II or III (regions defined in Figure \ref{fig:sensorLoc}).}
    \label{fig:wave_3y}
\end{figure}

Figure \ref{fig:wave_3y}a shows DDWD bases for three wall-normal positions (one for each region defined in Figure \ref{fig:sensorLoc}): $y^+=350$, $y/R=0.10$, and $y/R=0.59$. The wavelets are shown as functions of time scaled by the eddy turnover time (using the mean streamwise velocity at $y^+=350$). We plot in time instead of space to avoid the slight dilation of the wavelets that would occur from using Taylor's hypothesis. For all wall-normal positions, stage is synonymous with scale for the DDWD wavelets. 
For each wall-normal position, Figure \ref{fig:wave_3y}b quantifies the self-similarity of the wavelets across adjacent stages. Our observation of strong streamwise self-similarity at $y^+ = 350$ extends to the other wall-normal positions in the range of $l \in [2, 11]$. Figure \ref{fig:wave_3y}c quantifies the self-similarity of the DDWD wavelets across wall-normal positions by plotting the inner product in~\eqref{eq:alpha_fixedl} computed between the wavelets at $y^+ = 350$ and at the other wall-normal positions. A high degree of wall-normal self-similarity is present for stages $l \in [1, 11]$. 

\begin{figure}
    \centering
    \includegraphics[width=0.7\textwidth]{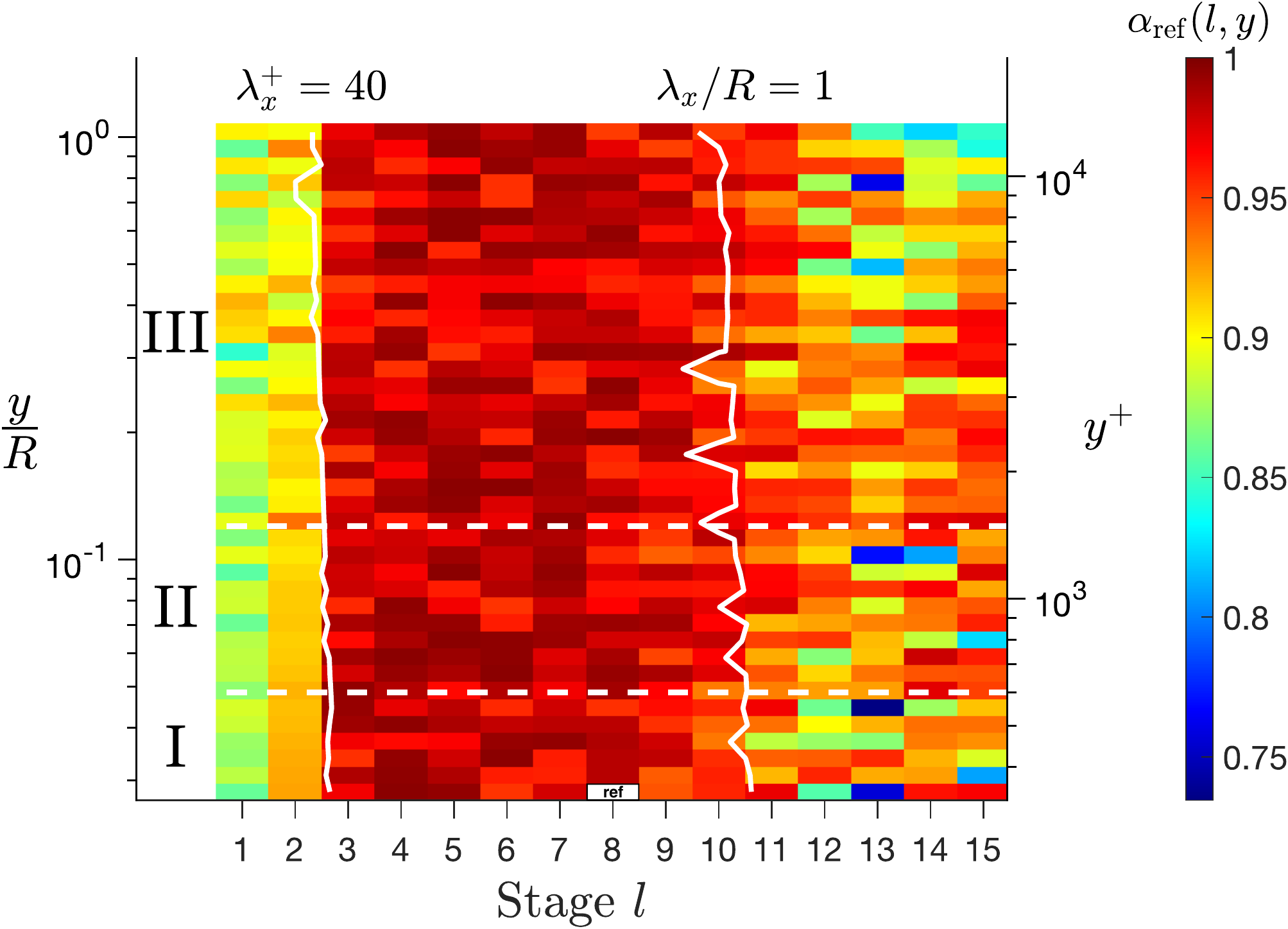}
    \caption{Self-similarity between the reference wavelet at $l_\text{ref}=8$ and $y_\text{ref}^+=350$ and all other wavelets. Dashed horizontal lines separate the wall-normal regions defined in Figure \ref{fig:sensorLoc}. Solid white lines are isolines of the streamwise length scale $\lambda_x$ of the wavelets.}
    \label{fig:alpha_l_y}
\end{figure}

Finally, we quantify the simultaneous self-similarity of the DDWD wavelets in the streamwise and wall-normal directions. To do so, we first choose a reference wavelet at stage $l_\text{ref} = 8$ and wall-normal position $y_\text{ref}^+=350$. Then, we calculate the inner product between this reference wavelet and all spatially localized wavelets ($l \leq 15$) at all wall-normal positions. We denote this inner product as $\alpha_\text{ref}(l,y)$. Note that either the reference wavelet or the wavelet at $(l,y)$ is scaled up to stage $\max(l,l_\text{ref})$ using the scaling operator $S$ before computing $\alpha_\text{ref}(l,y)$ (which is done in time, not space, to avoid slight dilations resulting from Taylor's hypothesis). Figure \ref{fig:alpha_l_y} plots $\alpha_\text{ref}(l,y)$.
\noindent
In Section \ref{sec:self_sim}, the DDWD wavelets at $y^+=350$ exhibited strong self-similarity for $l \in [2,11]$; however, based on the criterion $\alpha_\text{ref}\gtrsim 0.95$, Figure \ref{fig:alpha_l_y} suggests a range of $l \in [3,10]$ for all wall-normal positions. The slight discrepancy in the range of $l$ is caused by the slight difference between $\alpha_{l_a,l_b}$ and $\alpha_\text{ref}(l,y)$; we will use the latter range of $l \in [3,10]$ henceforth. Figure \ref{fig:alpha_l_y} suggests that DDWD wavelets, which form spatially localized, energy-containing motions, are self-similar between streamwise length scales of $\lambda_x^+ = 40$ to $\lambda_x/R = 1$ and between wall-normal length scales of $y^+=350$ to $y/R = 1$. Here, the self-similarity of the data has been assessed by the self-similarity of the DDWD basis. In the next section, a (perhaps) more direct method is explored where we assess the self-similarity of the data projected onto DDWD subspaces of different scales.

\subsection{Self-similarity of wavelet projections}
\label{sec:wave_proj}
The stage-$l$ wavelet projection $P_l z_i$ denotes the projection of $z_i \in \mathbb R^N$ onto the detail subspace $W_{-l}$ and is given by
\begin{equation}
    \label{eq:proj_dwt}
    P_l z_i = \sum_{m=0}^{N/2^l-1} \left\langle R^{2^l m}(\psi_{-l}), z_i \right\rangle R^{2^l m}(\psi_{-l}).
\end{equation}
Henceforth, $P_l u$ denotes the stage-$l$ wavelet projection of the velocity signal $u$ (or rather its segments) onto the DDWD subspaces. 

\begin{figure}
    \centering
    \includegraphics[width=0.6\textwidth]{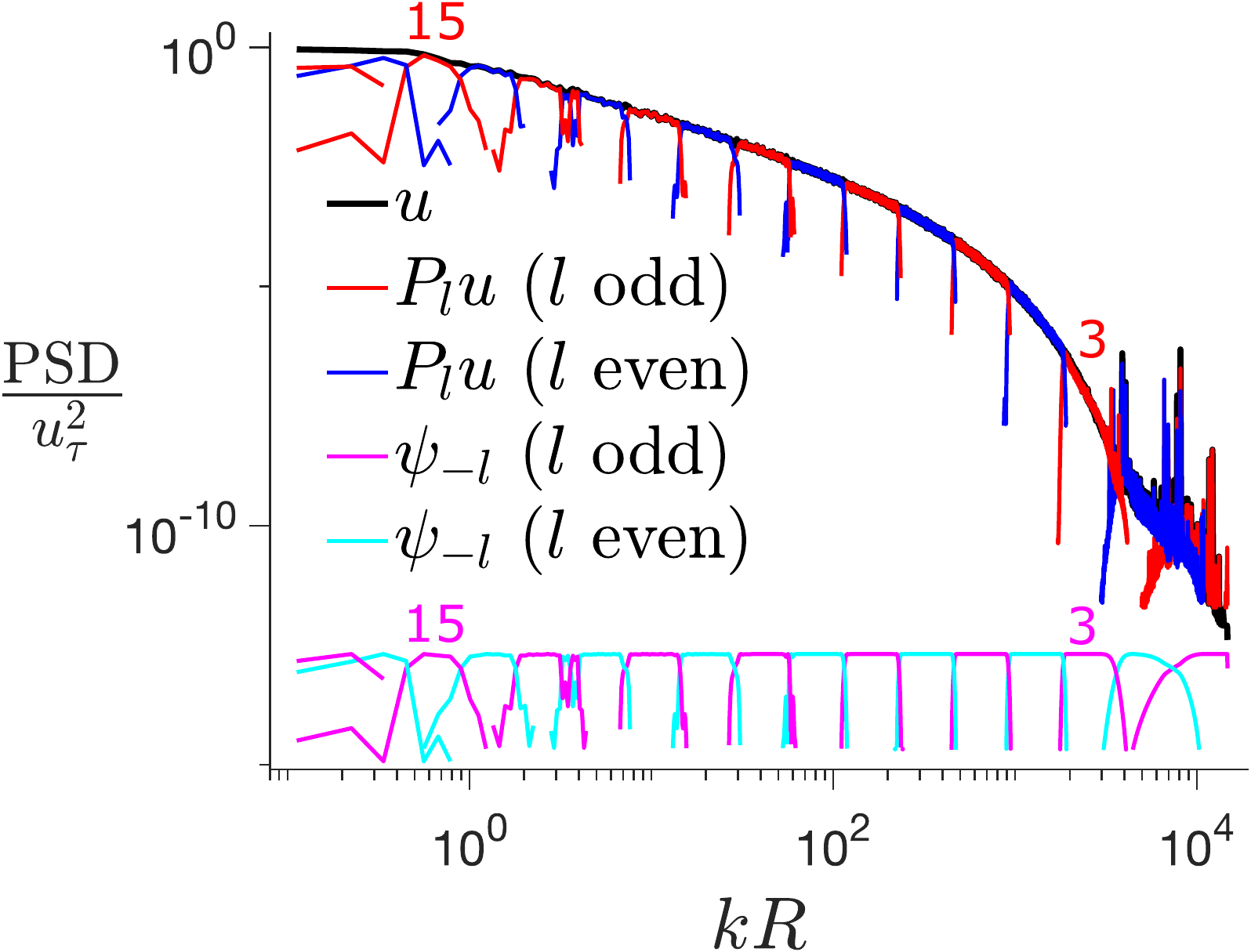}
    \caption{PSDs of $u$, its stage-$l$ DDWD wavelet projections $P_l u$, and the DDWD wavelets $\psi_{-l}$ at $y^+ = 350$. Note that the PSDs of the wavelets are manually shifted to be at the same height.}
    \label{fig:wave_proj_all_l}
\end{figure}

Wavelet projections, just like wavelets themselves, are multiscale. Larger-scale wavelet projections occupy the smaller-$k$ range, and vice versa. For $y^+=350$, Figure \ref{fig:wave_proj_all_l} shows the PSDs of $u$, $P_l u$, and the wavelets themselves, $\psi_{-l}$. \revise{As an aside, Figure \ref{fig:wave_proj_all_l} provides another potential reason why DDWD uncovers Meyer-like wavelets: the lower spatial localization (and thus higher spectral localization) of Meyer-like wavelets allows a narrower range of wavenumbers to be removed from the broadband velocity signal during the energy optimization.}

\begin{figure}
    \centering
    \includegraphics[width=0.8\textwidth]{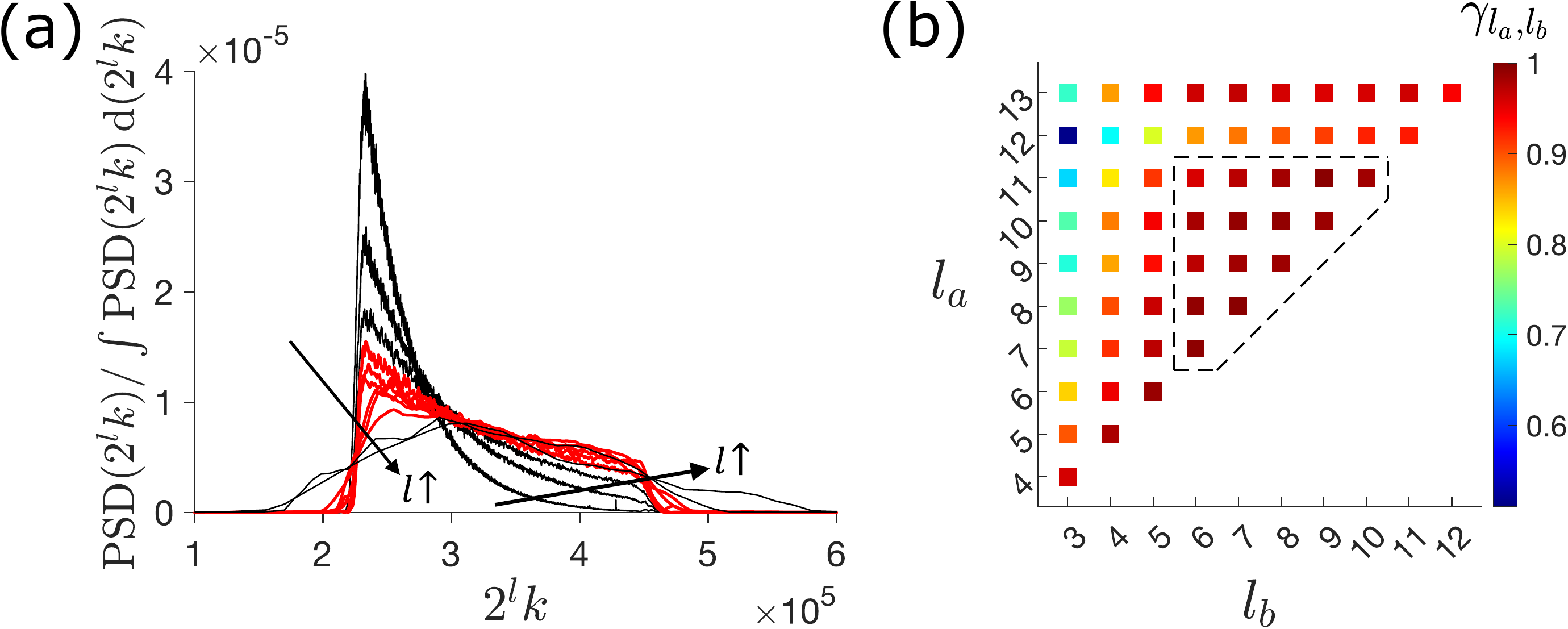}
    \caption{(a) Scaled and normalized PSDs of the DDWD wavelet projections in Figure \ref{fig:wave_proj_all_l} for $l \in [3,13]$. (b) Inner product between the curves in (a), where the dashed triangle indicates the region of high self-similarity ($\gamma_{l_a,l_b} \gtrsim 0.95$). Highly self-similar projections of scales $l \in [6,11]$ in (a) are colored red based on the triangular region in (b).}
    \label{fig:wave_proj_collapse}
\end{figure}

We assess the self-similarity of the wavelet projections in Figure \ref{fig:wave_proj_all_l} by comparing their PSDs. Figure \ref{fig:wave_proj_collapse}a shows the PSDs when scaled in $k$ by $2^l$ in order to account for the dyadic progression of scales; they are also normalized to have unit norm. The self-similarity of $P_l u$ is quantified by $\gamma_{l_a,l_b}$, which is the inner product between the scaled and normalized PSDs of scale $l_a$ and $l_b$; Figure \ref{fig:wave_proj_collapse}b plots this inner product. Just as with $\alpha$, we take $\gamma \gtrsim 0.95$ to indicate a high degree of self-similarity. Unlike DDWD wavelets, the projections are not highly self-similar until a much later stage of $l=6$. 

\begin{figure}
    \centering
    \includegraphics[width=1\textwidth]{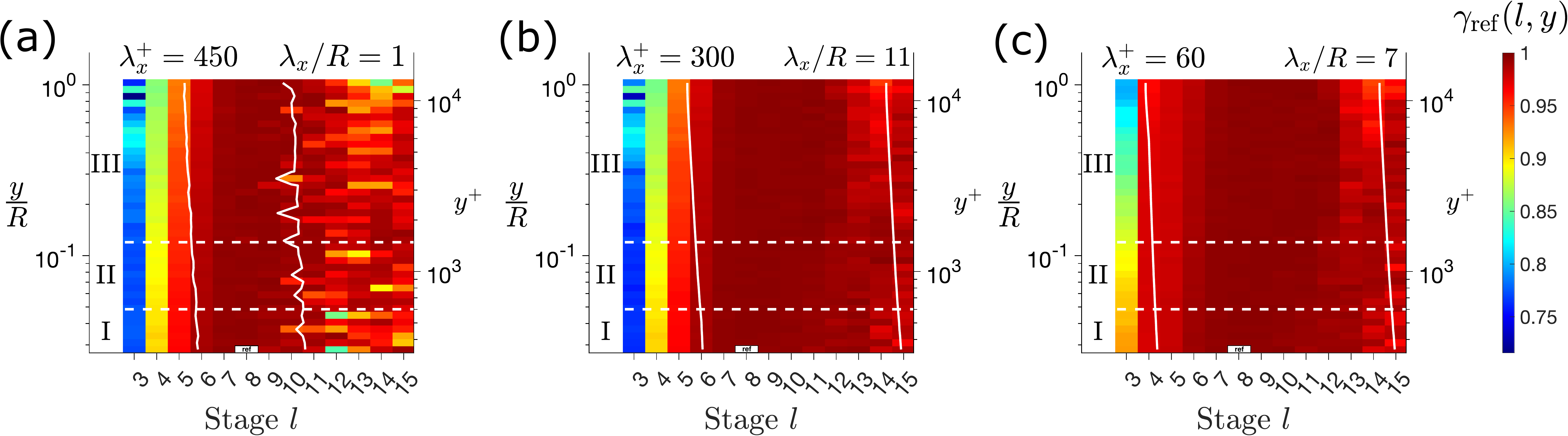}
    \caption{Same as Figure \ref{fig:alpha_l_y}, except with the inner product $\gamma$, which quantifies the self-similarity of the PSDs of wavelet projections using the (a) DDWD, (b) Meyer, and (c) db2 basis. \revise{Solid white lines are isolines of the streamwise length scale $\lambda_x$ of the wavelets.}}
    \label{fig:wave_proj_ip_arr_3_bases}
\end{figure}

Similar to the approach used in Section \ref{sec:sim_y}, we calculate the self-similarity of wavelet projections simultaneously in stage and wall-normal position by choosing a reference projection. We choose $l_\text{ref} = 8$ at the wall-normal position $y_\text{ref}^+=350$ again, and denote this inner product as $\gamma_\text{ref}(l,y)$; note that this inner product is done in frequency, not wavenumber, to avoid slight dilations resulting from Taylor's hypothesis. Figure \ref{fig:wave_proj_ip_arr_3_bases} plots $\gamma_\text{ref}(l,y)$ for the DDWD, Meyer, and db2 bases.
DDWD wavelet projections are self-similar in $y$ from $y^+=350$ to $y/R = 1$ and self-similar in $x$ (across stages) up until wavelets of wavelength $\lambda_x/R = 1$, exactly like DDWD wavelets (see Section \ref{sec:sim_y}). The only difference, however, is the lower bound of self-similarity in $x$---DDWD wavelets are self-similar starting from $\lambda_x^+ = 40$, while the projections are self-similar starting from $\lambda_x^+ = 450$. The reason for the difference is that by comparing the spectra of the projections, we simultaneously account for the shapes of the wavelets and the spectral content of the data, which yields a more stringent criterion for self-similarity. Figures \ref{fig:wave_proj_ip_arr_3_bases}b and \ref{fig:wave_proj_ip_arr_3_bases}c show that projecting the data onto traditional wavelet bases (with pre-defined and exactly self-similar shapes across scales) leads to overprediction of $\gamma_\text{ref}$ in $x$ by several stages since only spectral content is taken into account\revise{: $\gamma_\text{ref} \gtrsim 0.95$ for $l \ge 6$ ($\lambda_x^+ \approx 300$) when using Meyer wavelets and for $l \ge 4$ ($\lambda_x^+ \approx 60$) when using db2 wavelets. So even though Meyer wavelets may closely resemble DDWD wavelets (see Figure \ref{fig:wave_compare}), this result shows one advantage to using the DDWD basis rather than a prescribed wavelet basis for performing tasks such as projections.}

We now tie the results back to the AEH, which states that eddies in the log layer grow in size with distance from the wall and are self-similar. The energy-optimality principle of DDWD allows the shape of the DDWD wavelets and the DDWD wavelet projections to be each used as a criterion for measuring the self-similarity of the velocity. \revise{Our method defines self-similarity somewhat indirectly--- if two quantities (e.g., wavelets) across different stages are strongly similar (i.e., have an inner product value $\gtrsim 0.95$), we say they are self-similar.} The critical (and often-made) assumption is that self-similar fluctuating velocity signatures arise from self-similar eddies. In $x$, DDWD shows that eddies have a high degree of self-similarity from scales of $\lambda_x^+ = 40$ (based on the wavelets) or $\lambda_x^+ = 450$ (based on the wavelet projections) up to $\lambda_x/R = 1$, which is the streamwise extent of LSMs. VLSMs of streamwise extent $\lambda_x/R = \mathcal O(10 R)$ may also be self-similar, but to a lesser degree. In the wall-normal direction $y$, our results suggest that eddies are self-similar not just in the log layer but below and well above it. 

\revise{The self-similarity of the velocity in $x$ may correspond not only with attached eddies, but also with eddies associated with the energy cascade. In the next section, we determine the range of stages that accompany two spectral scaling laws: the $k^{-5/3}$ law related to the energy cascade, and the $k^{-1}$ law related (loosely) to self-similarity of wall-bounded turbulence.}

\subsection{Energy and spectral scaling laws}
\label{sec:energy}
The log law for the streamwise turbulence intensity, $\overline{u u}$, is a key feature of wall-bounded turbulence that has been widely observed in experiments and simulations \citep{Hultmark2012, Rosenberg2013, winkel2012turbulence, marusic2013logarithmic, lee2015direct}. There are three models that have derived this log law. The first is Townsend's AEH \citep{Townsend1976}, which was a statistical model that assumed the contributions to the turbulence intensity from attached eddies were self-similar. The second is the AEM \citep{Perry1982}, which was a physical model inspired by the AEH that assigns a shape to the attached eddies. The third is the spectral AEM \citep{Perry1986}, which relied on scaling and dimensional arguments to argue that the PSD of $u$ should exhibit a $k^{-1}$ power law in an intermediate range of wavenumbers; integrating the PSD gives a log law for $\overline{u u}$. So, the spectral AEM shows that a $k^{-1}$ power law in the PSD of $u$ is consistent with the existence of a log law for $\overline{u u}$. Transitively, one could then say that the existence of a $k^{-1}$ power law would be consistent with the existence of geometrically self-similar eddies proposed by the AEH.

Here, we attempt to search for a $k^{-1}$ scaling in wavelet space instead of $k$-space. As discussed in Section~\ref{sec:overview}, the energy contained in a signal is partitioned into the highlighted wavelet subspaces in Figure~\ref{fig:subspaces}. The mean energy of our dataset (segments of $u$) contained in subspace $l$ is
\begin{equation}
    \label{eq:WED}
    E_l = \frac{1}{M} \sum_{i=1}^M \| P_l z_i \|^2 = \frac{1}{M} \sum_{i=1}^M \sum_{m=0}^{N/2^l-1} \left\vert \left\langle R^{2^l m}(\psi_{-l}), z_i \right\rangle \right\vert ^2.
\end{equation}
For a signal of length $N=2^p$, $E_l$ is defined for stages $l \in [1,p]$ (the detail subspaces) and stage $p_a$ (the final approximation subspace). At stage $p_a$, $\psi_{-l}$ in~\eqref{eq:WED} is replaced by $\phi_{-p}$. We call the distribution of energy amongst these subspaces the wavelet energy distribution (WED). 

The key difference between the PSD and the WED is that the PSD gives the energy per unit wavenumber, while the WED gives the energy contained in a range of wavenumbers. Assuming self-similar wavelets, $E_l$ gives the energy across a range of wavenumbers $\Delta k_l \sim 2^{-l}$ (see Figure~\ref{fig:wave_proj_all_l}). Therefore a quantity analogous to the PSD, which we call the wavelet power spectrum (WPS), is $\tilde E_l = E_l/\Delta k_l \sim 2^l E_l$, as used by \cite{Meneveau1991}. Meneveau's definition of the WPS was carefully normalized to match the energy contained in the continuous signal. The normalization does not change how $E_l$ scales with $l$, which is our main interest.

We now discuss the equivalence of power laws between the PSD, WED, and WPS. Since the wavelength $\lambda_x$ of a wavelet is proportional to both $2^l$ and $1/k$ ($k$ being the central wavenumber of the wavelet), a power law of $\phi_{uu}(k) \sim k^{-\gamma}$ in the PSD will appear either as $\tilde E_l \sim (1/\lambda_x)^{-\gamma} \sim (1/2^l)^{-\gamma} \sim 2^{\gamma l}$ or $E_l \sim 2^{(\gamma-1) l}$. The power law scaling for the WED can also be derived by integrating a power law for the PSD across wavenumbers corresponding to adjacent stages:
\begin{equation}
    E_l \sim \int_{2^{-l}}^{2^{-(l+1)}} k^{-\gamma} \,\text{d} k = \frac{2^{1-\gamma} - 1}{1 - \gamma} 2^{(\gamma-1) l}.
    \label{eq:WED_power_law}
\end{equation}
Henceforth, we work with the WED. Therefore \revise{$\gamma=5/3$} and \revise{$\gamma=1$} lead to $E_l\sim 2^{2l/3}$ and $E_l\sim 2^0\sim\textrm{constant}$, respectively. 

\begin{figure}
    \centering
    \includegraphics[width=1\textwidth]{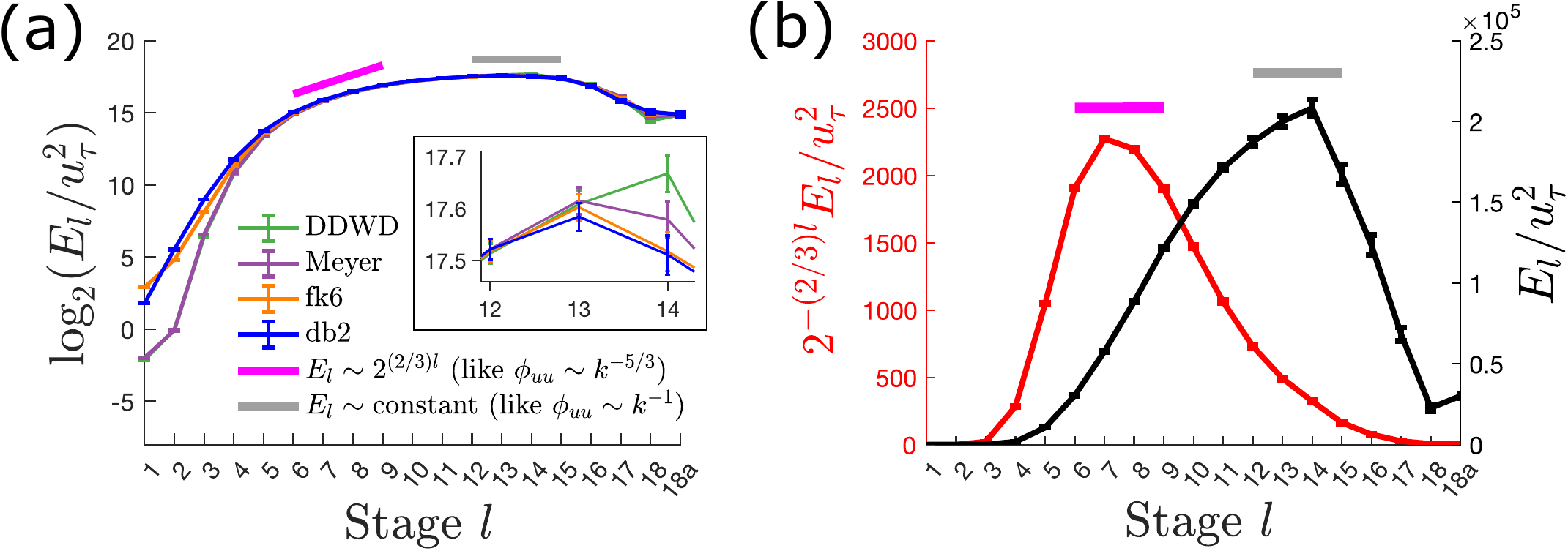}
    \caption{(a) Wavelet energy distribution (WED) of the signal measured at $y^+=350$ calculated using the DDWD basis and 3 known wavelet bases (the green and purple curves overlap). (b) Two versions of the DDWD WED from (a), one of which is premultiplied. Note, error bars are standard errors.}
    \label{fig:WED_yloc1}
\end{figure}

As before, we first discuss the WED results for the velocity signal measured closest to the wall. Figure \ref{fig:WED_yloc1}a shows the WED for this signal when $E_l$ is computed with the DDWD basis and three traditional wavelet bases. As seen in Figure \ref{fig:WED_yloc1}a and its inset, $E_l$ is lower in the DDWD basis until stage $l=14$, which is the most energetic (but not the largest) stage. The reader will note that the WED at stage 18a (which corresponds to the squared mean) is non-zero despite $u$ having  zero mean. The reason is that, like the PSD, the WED is estimated from an ensemble of segments of $u$ which may themselves have non-zero mean. 
By appropriately multiplying the WED of the DDWD basis, the two power laws in Figure \ref{fig:WED_yloc1}a appear as plateau regions in Figure \ref{fig:WED_yloc1}b. Note that the WED plotted on the right vertical axis in Figure \ref{fig:WED_yloc1}b is the same as that in Figure \ref{fig:WED_yloc1}a, except plotted with a linear scale for the vertical axis to better discern whether a plateau region exists. A plateau region corresponding to a $k^{-5/3}$ scaling is not significant, which is common for measurements close to the wall \citep{Vallikivi2015,Rosenberg2013}. \revise{Nonetheless, one can still say that the inertial subrange occurs around stages 6 through 9, and that the self-similarity results found previously for these stages may also be attributed to the energy cascade and not solely to the energy-containing motions from Townsend's AEH.} A plateau region corresponding to a $k^{-1}$ scaling is even less apparent.

\begin{figure}
    \centering
    \includegraphics[width=0.6\textwidth]{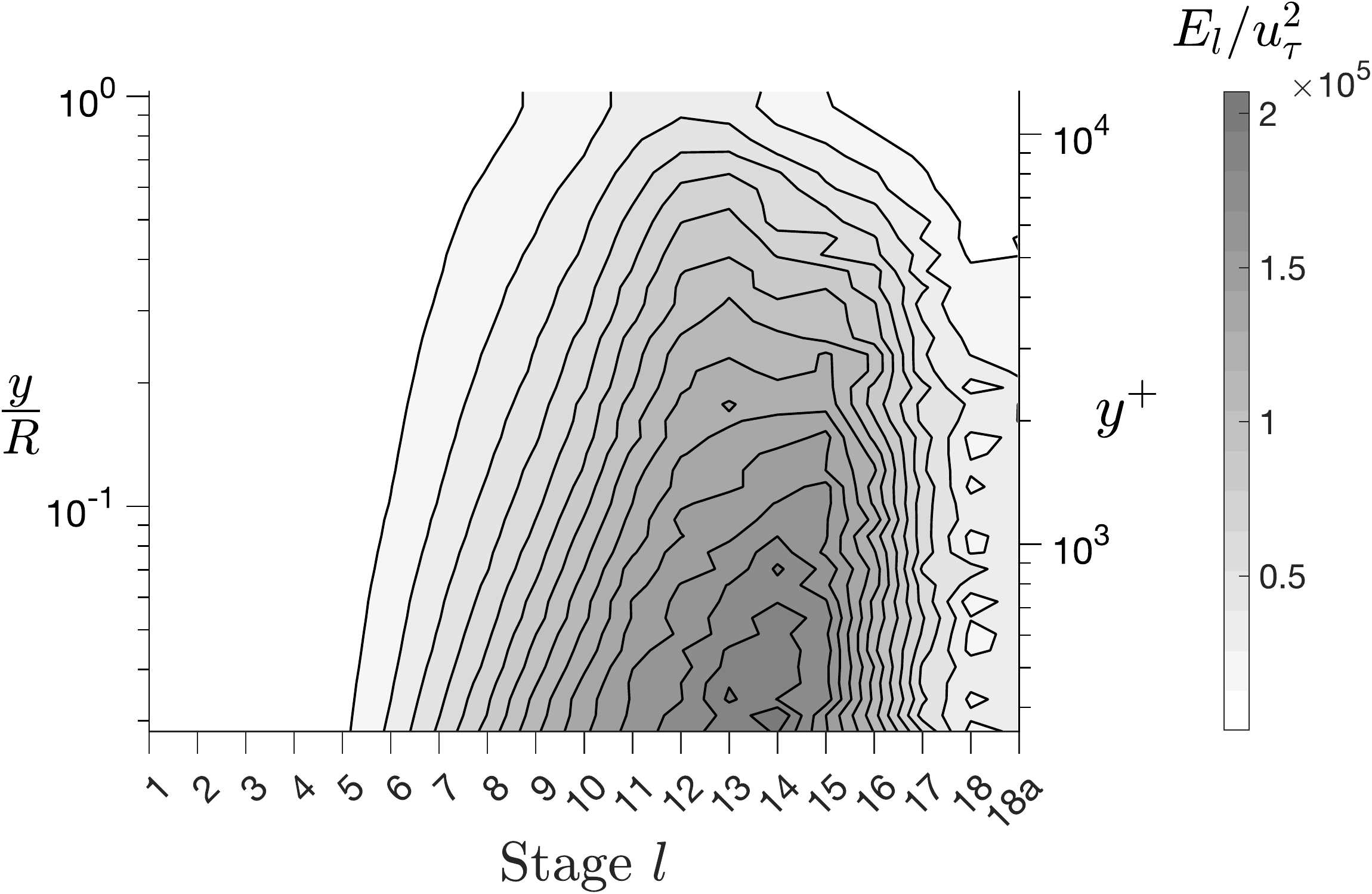}
    \caption{(a) Wavelet energy distribution (WED), calculated using the DDWD basis, at all wall-normal positions. This plot is analogous to a premultiplied PSD spectrogram.}
    \label{fig:WED_arr}
\end{figure}

Figure \ref{fig:WED_arr} shows $E_l$ (calculated with the DDWD basis) at all wall-normal positions. This plot is analogous to a premultiplied PSD spectrogram, where a $k^{-1}$ scaling would manifest as a flat region. \revise{There is a relatively flat region for $y^+\lesssim 1000$  and $12\leq l \leq 15$. In the premultiplied PSD spectrograms of other studies, a flat region appears at a similar location and is referred to as the ``outer peak'' \citep{hutchins_marusic_2007,Hultmark2012}. In \cite{hutchins_marusic_2007}, they propose that the outer peak corresponds to the presence of VLSMs, which is consistent with our data since $l \approx 13$ corresponds to $\lambda_x = \mathcal O(10 R)$ (see Figure \ref{fig:self_sim_x}a). The spectral AEM \citep{Perry1986} states that the $k^{-1}$ law associated with attached eddies exists at length scales in the overlap of the $\mathcal O(y)$ region (stages 6 through 11 here) and the $\mathcal O(R)$ region (stages 11 through 15 here)---therefore, although a somewhat flat region appears at $l \approx 13$ in Figure \ref{fig:WED_arr}, it occurs at length scales that are too large to be considered a true $k^{-1}$ law that is associated with attached eddies.}

A $k^{-1}$ power law is rarely observed in experiments and simulations \citep{Nickels2005,Chung2015,Rosenberg2013}. One possible reason for the frequent absence of a $k^{-1}$ power law is that it is only expected to appear when there is enough scale separation in the flow, which occurs at high friction Reynolds numbers. \cite{Marusic2019} estimate that $\text{Re}_\tau = \mathcal O(10^6)$ is required to observe even a small region with $k^{-1}$ scaling, which is not achievable except in the atmospheric boundary layer. Since our Superpipe data has $\text{Re}_\tau = \mathcal O(10^4)$, a $k^{-1}$ power law is not expected to appear clearly. Moreover, \cite{Hwang2022} showed that even without the presence of a $k^{-1}$ power law, the framework of the spectral AEM can still successfully yield a log law for $\overline{u u}$, which is a key feature of wall-bounded turbulence. Therefore, within the context of the spectral AEM, the existence of a $k^{-1}$ power law may not be a necessary feature of wall-bounded turbulence. As for whether geometric self-similarity is a feature of eddies in wall-bounded turbulence, a consensus has not been reached. In the next section, we compare the self-similarity results found by DDWD to the results of other methods.

\section{Comparison of ranges of self-similarity}
\label{sec:lit_compare}

Several turbulent structures have been found and studied in the literature, many of them displaying self-similarity across a range of scales. Here, we compare the ranges of self-similarity for these structures and for the localized structures found herein via DDWD. 

Hairpin vortices, first proposed by \cite{Theodorsen1952}, may be attached eddies. These coherent structures are theorized to be generated in a bottom-up process \citep{Zhou1999,Tomkins2003,adrian2007hairpin}. At the smallest scale, hairpin vortices of height $\mathcal O(100 \delta_\nu)$, where $\delta_\nu$ is a wall unit, are generated in the viscous layer. Then, these hairpin vortices merge with each other in a pairing process to give larger hairpin vortices, which in turn form packets of hairpin vortices (e.g., LSMs and VLSMs) with height $\sim 0.3 \delta$ to $1 \delta$, where $\delta$ is the turbulent boundary layer (TBL) height. However, this bottom-up generation process is not universally agreed upon. \cite{Flores2006} showed that the structure and dynamics of LSMs and VLSMs in the outer region were not affected upon suppressing the generation of hairpin vortices in the viscous layer. Furthermore, \cite{Zhou2022} used correlation methods to show that the accumulation of streaks in the buffer layer is not necessary for the generation of LSMs. On the other hand, a top-down process was found in which LSMs induced the spanwise drift of streaks.

Another candidate for attached eddies is the structures related to the SSP \citep{Hamilton1995,Waleffe2001}, namely quasi-streamwise vortices and streaks. Although SSP structures are generally considered to exist only in the buffer layer, \cite{Eckhardt2018} found exact coherent states (ECS; see \cite{kawahara2012significance} and \cite{graham2021exact}), corresponding to SSP structures, that exist in the outer region and are self-similar. Further support of SSP structures being self-similar is found in \cite{Hwang2015} and \cite{Cossu2017}, where overdamped large-eddy simulations (LES) were used to isolate structures at a specific scale. In these simulations, the SSP structures representing attached eddies were found to be self-similar; note that an ``attached eddy'' in this case is composed of the dynamically interconnected quasi-streamwise vortices and streaks. The largest self-similar attached eddy was found to be comprised of a streaky motion of streamwise extent $15 h$ ($h$ is the channel half-height) and height $\sim 0.15 h$ that is surrounded by quasi-streamwise vortices of streamwise extent $3 \sim 4 h$ and height $0.75 h \sim 1 h$. This aspect ratio of the self-similar SSP structures is maintained throughout the outer region all the way down to the viscous layer at $\mathcal O(10 \delta_\nu)$. The largest streak component resembles a VLSM and the largest quasi-streamwise vortices component resembles LSMs. Furthermore, the overdamped LES results imply that turbulence can be sustained at each individual scale without any forcing from smaller scales, thus lending more support against the bottom-up generation of vortices discussed previously. 

An operator-based method called resolvent analysis was shown to produce reconstructed velocity fields containing self-similar, wall-attached structures \citep{Hwang2010,McKeon2017,Moarref2013}. These structures are confined to the log region and interestingly show a quadratic scaling of streamwise extent with distance from the wall instead of a linear scaling as proposed in the AEH.

A method to isolate coherent structures that are physically attached to the wall is spectral coherence analysis (SCA) \citep{Baars2017,Baars2020}. SCA uses the magnitude of the cross-spectrum of a two-point streamwise fluctuating velocity signal (one measured at the wall, and one measured in the log layer) to measure the coherence of structures with the wall at each scale $1/k$. By arguing that a self-similar hierarchy of attached structures manifests in a certain pattern in the cross-spectrogram, it was shown that a TBL flow contained self-similar attached structures for heights $80 \delta_\nu$ to $0.71 \delta$ and streamwise wavelengths $\mathcal O(1000 \delta_\nu)$ to $\sim 10 \delta$. These extents were found to be universal for a friction Reynolds number range of $\mathcal O(10^3)$ to $\mathcal O(10^6)$. The aspect ratio and range of attached-eddy heights found using SCA correspond closely to those of the self-similar SSP structures found using overdamped LES in \cite{Hwang2015} and \cite{Cossu2017}.

Modal decomposition techniques, like DDWD, aim to represent coherent structures with modes. In \cite{Hellstrom2016}, the radial POD modes of pipe flow were found to be self-similar for heights $~150 \delta_\nu$ to $~0.6 R$. In \cite{Wang2022}, three-dimensional POD modes of a turbulent channel flow were decomposed into individual eddies called ``POD eddies;'' the wall-attached POD eddies were found to be physically self-similar for heights $\mathcal O(100 \delta_\nu)$ to $0.4 h$. In \cite{He2019}, a wavelet decomposition in time (using db20 wavelets) and POD in space were performed on TBL flow, which yielded modes that were self-similar up to height $\sim 0.5 \delta$.

\revise{Table \ref{tab:compare} compares the ranges of self-similarity found by the methods discussed above to the ranges DDWD finds.} Compared to the studies above, our DDWD results suggest similar ranges of self-similarity of eddies: streamwise extent starting from $40 \delta_\nu$ (based on the wavelets) or $450 \delta_\nu$ (based on the wavelet projections) up to $1R$ (and possibly $\mathcal O(10 R)$), and heights starting from $350 \delta_\nu$ to $1R$. \revise{Since the closest measurement of the velocity was at 350 wall units, the true lower bound for the self-similarity in the wall-normal direction of the flow experiment may be smaller like in other studies. Finally, we note that while the DDWD wavelet projections give a lower bound of streamwise self-similarity that matches well with other studies, the same cannot be said about DDWD wavelets---one reason could be that DDWD wavelets are uncovering self-similarity of coherent structures related to the energy cascade, rather than the AEH, at these smaller scales.}

\begin{table}
    \begin{center}
    \begin{tabular}{p{0.22\linewidth}                                                        p{0.1\linewidth}     p{0.09\linewidth}   p{0.09\linewidth}     p{0.09\linewidth}   p{0.14\linewidth}        p{0.08\linewidth} p{0.05\linewidth} }
        \centering\textbf{Method}                                                            & $x_1 / \delta_\nu$ & $x_2 / \delta_o$ & $y_1 / \delta_\nu$ & $y_2 / \delta_o$ & Re$_\tau$              & Type            & Data             \\ \hline
        \centering\textbf{Flow visualization} \citep{Zhou1999,Tomkins2003,adrian2007hairpin} & varies             & varies           & $\mathcal O(100)$  & $\sim$0.3--1     & varies                 & varies          & varies           \\
        \centering\textbf{Overdamped LES: streaks} \citep{Hwang2015}                         & 1000               & 15               & 10                 & 0.15             & 1000--2000             & channel         & sim.             \\
        \centering\textbf{Overdamped LES: vortices} \citep{Hwang2015}                        & 200--300           & 3--4             & 50--70             & 0.75--1          & 1000--2000             & channel         & sim.             \\
        \centering\textbf{Spectral coherence analysis} \citep{Baars2017}                     & 1120               & 10               & 80                 & 0.71             & 2000--$1.4\times 10^6$ & TBL             & varies           \\
        \centering\textbf{POD} \citep{Hellstrom2016}                                         & n/a                & n/a              & $\sim$150          & $\sim$0.6        & 1300-2500              & pipe            & exp.             \\
        \centering\textbf{POD eddies} \citep{Wang2022}                                       & $\sim$600          & $\sim$1.5        & $\sim$130          & $\sim$0.4        & 2000                   & channel         & sim.             \\
        \centering\textbf{POD + wavelet transform} \citep{He2019}                            & $\sim$200          & $\sim$2          & $\sim$50           & $\sim$0.5        & 235                    & TBL             & exp.             \\
        \centering\textbf{DDWD: wavelets}                                                    & 40                 & $\sim$1          & 350                & 1                & 12400                  & pipe            & exp.             \\
        \centering\textbf{DDWD: wavelet projections}                                         & 450                & $\sim$1          & 350                & 1                & 12400                  & pipe            & exp.             \\
    \end{tabular}
    \caption{Comparison of DDWD to other methods in estimating the range of streamwise self-similarity, $[x_1,x_2]$, and wall-normal self-similarity, $[y_1,y_2]$, of attached eddies in simulation (sim.) and experimental (exp.) datasets of wall-bounded turbulence. $\delta_\nu$ is a wall-unit, and $\delta_o$ is the outer-length scale.}
    \label{tab:compare}
    \end{center}
\end{table}

\section{Conclusions}
\label{sec:conc}

DDWD was used to extract energetic, spatially localized structures from experimental measurements of the streamwise velocity in a fully developed turbulent pipe flow in the Princeton Superpipe. The structures---which we interpret as being reflective of underlying eddies---were found to be self-similar across streamwise extents of 40 wall units to one pipe radius, and across wall-normal positions between $y^+ = 350$ to $y/R = 1$. We found that the structures were similar in shape to Meyer wavelets. An analysis of the spectra of projections of the data onto DDWD subspaces yielded the same bounds for self-similarity except with streamwise self-similarity starting at a larger length scale of 450 wall units. This wavelet projection metric might be more accurate since it considers both the shape of the wavelet (learned from data) and the Fourier content of the data\revise{, and since some studies \citep{Hwang2015,Baars2017} propose the lower bound of streamwise self-similarity starting at $\mathcal O(100)\text{--}\mathcal O(1000)$ wall units}.

While our bounds for streamwise self-similarity are consistent with the literature, the large range of wall-normal self-similarity seems contrary to the AEH \citep{Townsend1976}, which states that attached eddies are self-similar only in the log layer; however, self-similarity of structures well beyond the log layer has been observed in other studies as well \citep{Hwang2015,Baars2017,Hellstrom2016}.

\revise{Characterizing attached eddies is an important first step in gaining mechanistic insight for how they are generated and sustained. In our previous work with HIT data \citep{Floryan2021}, DDWD returned self-similar, Meyer-like wavelets for stages corresponding to the inertial subrange. For the Superpipe data, the inertial subrange corresponds to stages 6 through 9 (450 wall units to $\sim$0.3 pipe radii); however, here we see self-similar, Meyer-like wavelets starting from stage 3 and ending at stage 10 or 11 (note, wavelet length scales increase by a factor of $\sim$2 with stage). While it is interesting that DDWD returns Meyer-like wavelets for two different types of turbulent flows, two caveats should be made. First, for both datasets, our 1D wavelets can only speak to 1D self-similarity of structures, and thus cannot provide detailed information about 3D structure (e.g., degree of isotropy). Second, for the Superpipe data, DDWD cannot differentiate between velocity structures induced by anisotropic attached eddies and those induced by isotropic eddies corresponding to the energy cascade---especially since the mechanisms governing these two types of eddies are likely coupled and complex.}

DDWD wavelets are a natural choice for extracting spatially localized, multiscale velocity structures (induced by eddies) from data. In search for eddies that are physically attached to the wall, future efforts will be directed toward applying DDWD to multi-dimensional data and developing a DDWD analog of the spectral wall-coherence method given in \cite{Baars2017,Baars2020}.

\begin{acknowledgments}
This work was supported by AFOSR FA9550-18-1-0174 and ONR N00014-18-1-2865 (Vannevar Bush Faculty Fellowship). The authors are grateful to Marcus Hultmark and Matthew K. Fu of Princeton University for use of the Superpipe data and helpful discussions. The authors additionally thank Kelly Y. Huang, Gabriel G. Katul, and Alexander J. Smits for helpful discussions. 
\end{acknowledgments}

Declaration of Interests. The authors report no conflict of interest.

\revise{Data availability statement. The code that supports the findings of this study is openly available at \href{https://github.com/dfloryan/DDWD}{https://github.com/dfloryan/DDWD}.}

\appendix

\section{Additional details about use and validation of DDWD}
\label{sec:appendix}

\subsection{Details of obtaining an 18-stage DDWD basis}
\label{sec:app_18s}
As mentioned in Section \ref{sec:scales}, one iteration of the optimization problem has a computational complexity of $\mathcal{O}(N^2 \log_2 N + N^2 M)$, and the number of iterations required for the BFGS algorithm to converge grows with $N$. The optimization problem is intractable for $N=2^{18}$. To make the problem tractable, we use the fact that wavelets are spatially localized. 

The principal idea behind drastically reducing the computation time of DDWD is to find shortened wavelets of length $N_s \ll N$, and then pad them with zeros until they have length $N$. As long as a wavelet of length $N$ is localized within a region of length less than $N_s$, it can be accurately represented by a wavelet of length $N_s$. Reciprocally, a wavelet of length $N_s$ that is localized can be extended to a wavelet of length $N$ by padding it with zeros. Since the spatial extent of a wavelet increases with the stage $l$ roughly as $2^l$, our idea will seemingly only work for the early stages of the wavelet hierarchy corresponding to small-scale wavelets. We circumnavigate this issue by working with filters (also called wavelet generators) instead of directly working with wavelets. In what follows, we provide a minimal background on how filters are used to produce a wavelet basis (further details appear in \cite{Floryan2021}), and how we modify the approach to create an optimization problem that is tractable for large $N$. 

We work in the space $\mathbb{R}^N$. A first-stage wavelet basis for $\mathbb{R}^N$ is an orthonormal basis of the form 
\begin{equation}
    \label{eq:1s_wavelet_basis}
    \{ R^{2m} u_1 \}_{m=0}^{N/2-1} \cup \{ R^{2m} v_1 \}_{m=0}^{N/2-1}.
\end{equation}
That is, $\mathbb{R}^N$ is spanned by $N/2$ mutually orthonormal even translates of $u_1, v_1 \in \mathbb{R}^N$. The two sets in~\eqref{eq:1s_wavelet_basis} are respectively bases for the subspaces $V_{-1}$ and $W_{-1}$ that appear in Figure~\ref{fig:subspaces}. The vectors $u_1$ and $v_1$ are called filters, with $u_1$ being equal to the father wavelet $\phi_{-1}$ introduced in Section~\ref{sec:overview}, and $v_1$ being equal to the mother wavelet $\psi_{-1}$. A key point is that wavelets can be constructed from the filters, and working directly with the filters confers computational advantages. 

The orthonormality conditions for $v_1$ and its even translates can be stated in Fourier space as
\begin{equation}
    \label{eq:orthog_cond_spectral}
    \left| \hat v_1(n) \right|^2 + \left| \hat v_1(n + N/2) \right|^2 = 2 \, , \quad n = 0, 1, \dots, N/2-1,
\end{equation}
where $\, \hat{} \,$ denotes the discrete Fourier transform (DFT). For a vector $z \in \mathbb{R}^N$, $z(n)$ is its $n^\text{th}$ component, indexed beginning from zero. The same conditions hold for $u_1$, and additional conditions are needed to ensure that the even translates of $u_1$ and $v_1$ are mutually orthonormal. Given a $v_1$ that satisfies~\eqref{eq:orthog_cond_spectral}, $u_1$ can be constructed from $v_1$ (and vice versa) so that~\eqref{eq:1s_wavelet_basis} is an orthonormal basis. 

For $N = 2^p$, one can construct a $p^\text{th}$-stage wavelet basis for $\mathbb{R}^N$ corresponding to the $p$ stages of the wavelet hierarchy in Figure~\ref{fig:subspaces}. To do so, we require $p$ sets of filters. Since the dimension of subspaces $V_{-l}$ and $W_{-l}$ is $N/2^{l}$, the filters become progressively shorter at later stages; specifically, $u_l, v_l \in \mathbb{R}^{N/2^{l-1}}$. The filter $v_l$ at stage $l$ must satisfy the orthonormality constraints in~\eqref{eq:orthog_cond_spectral} (with $N$ replaced by $N/2^{l-1}$), and $u_l$ can be constructed from $v_l$. 

Given the filters, the wavelets at stage $l$ are given by 
\begin{align}
  \label{eq:exwave}
  \psi_{-l} &= u_1 * U(u_2) * U^2(u_3) * \ldots * U^{l-2}(u_{l-1}) * U^{l-1}(v_l), \\
  \phi_{-l} &= u_1 * U(u_2) * U^2(u_3) * \ldots * U^{l-2}(u_{l-1}) * U^{l-1}(u_l),
\end{align}
where $*$ denotes convolution and $U$ is the upsampling operator. The upsampling operator doubles the length of its input and is defined by $U(z)(n) = z(n/2)$ for $n$ even and 0 for $n$ odd. The $N/2^l$ translates by $2^l$ of $\psi_{-l}$ and $\phi_{-l}$ span subspaces $W_{-l}$ and $V_{-l}$, respectively. Since the wavelets are constructed by repeated convolutions of the filters, localized filters yield localized wavelets. For traditional wavelets, the stage-$l$ wavelet has a fixed length $N$ and is localized over a length proportional to $2^l$. Conversely, the corresponding filters have varying lengths of $N/2^{l-1}$ but are localized over constant lengths. We will make use of this feature for our modified DDWD algorithm. 

In DDWD, we find optimal filters stage by stage. At the first stage, the filters have length $N$, making the optimization problem intractable for $N = 2^{18}$. At stage $l$, the filters have length $N/2^{l-1}$, making the optimization problem easier as we move to later stages. Thus for large $N$, only the initial stages are intractable, so only the initial stages need to be modified.

We now discuss the modified DDWD algorithm. Let $Z \in \mathbb R^{N \times M}$ contain an ensemble of $M$ data vectors of length $N$. The wavelets (and corresponding filters) at the first stage will be highly localized, permitting us to use data vectors of length $N_s \ll N$ to find them. In this work, we use $N_s = 2^{p_s} = 2^7$. The first step is to resize $Z$ into $Z_{s} \in \mathbb R^{N_s \times M N/N_s}$ so that $Z_{s}$ contains the shortened data vectors. Standard DDWD is then applied to $Z_{s}$ to solve for $v_{1s} \in \mathbb R^{N_s}$, which has computational cost $\mathcal{O}(N_s^2 \log N_s + N_s^2 (M N/N_s))$ per iteration of the BFGS solver. Moreover, the BFGS solver requires far fewer iterations to converge for a lower-dimensional problem. Then, the filter $v_{1s}$ is padded with zeros until it has length $N$ to give the full-length filter $v_1$. In general, padding yields a filter $v_1$ that is slightly non-orthogonal to its even translates. This is easily remedied by multiplying the Fourier coefficient pairs that appear in~\eqref{eq:orthog_cond_spectral} by the unique positive real number that will make condition~\eqref{eq:orthog_cond_spectral} hold, restoring the orthonormality property. Next, $v_1$ is used to project $Z$ onto the subspace $V_{-1}$ to form $Z_1 \in \mathbb R^{N/2 \times M}$. The coarsened data matrix $Z_1$ is used to calculate the filter at the next stage, $v_2$, the same way $Z$ is used to calculate $v_1$. Note that as we move on to the next stage, the dimension of the coarsened dataset is halved, making the optimization problem increasingly tractable. We repeat this process recursively until stage $l=p-p_s+1$, at which point the coarsened dataset has dimension $2^{p_s}$ without being shortened. From this stage onward, DDWD without modifications is performed. 

In the end, the modified algorithm returns the full $p$-stage DDWD basis for $\mathbb{R}^N$ but with vastly reduced computing time. The computational savings are enabled by only ever directly using data vectors of maximal length $N_s \ll N$. Crucially, presupposed localization of the wavelets allows us to use shortened data vectors without distorting the resulting wavelets, and the validity of this presupposition can be checked after the fact (we have checked that $N_s = 2^7$ is indeed sufficient).

\subsection{DDWD validation using synthetic, self-similar signals}
\label{sec:app_synth}

To validate the results produced by DDWD, we test whether the algorithm can correctly recover the shape of a wavelet from a synthetic signal that is a superposition of a known wavelet across many scales and locations. 

We construct the synthetic signal so that it mimics the velocity measurement at $y^+ = 350$ in the Superpipe. The synthetic signal, $u_\text{synth}$, has length $2^{24}$. We generate $u_\text{synth}$ by specifying values for its wavelet coefficients and then perform an inverse discrete wavelet transform. The coefficients corresponding to each wavelet subspace are sampled from a normal distribution and then normalized such that the energy in each wavelet subspace matches the WED in Figure \ref{fig:WED_yloc1}a (we choose to match the WED of the Meyer wavelet in that figure). The energy in stages $l \in [19, 24]$ is set equal to zero. We create a dataset of vectors of length $2^{18}$ from this signal the same way that we created a dataset from the Superpipe velocity signal. 

\begin{figure}
    \centering
    \includegraphics[width=0.8\textwidth]{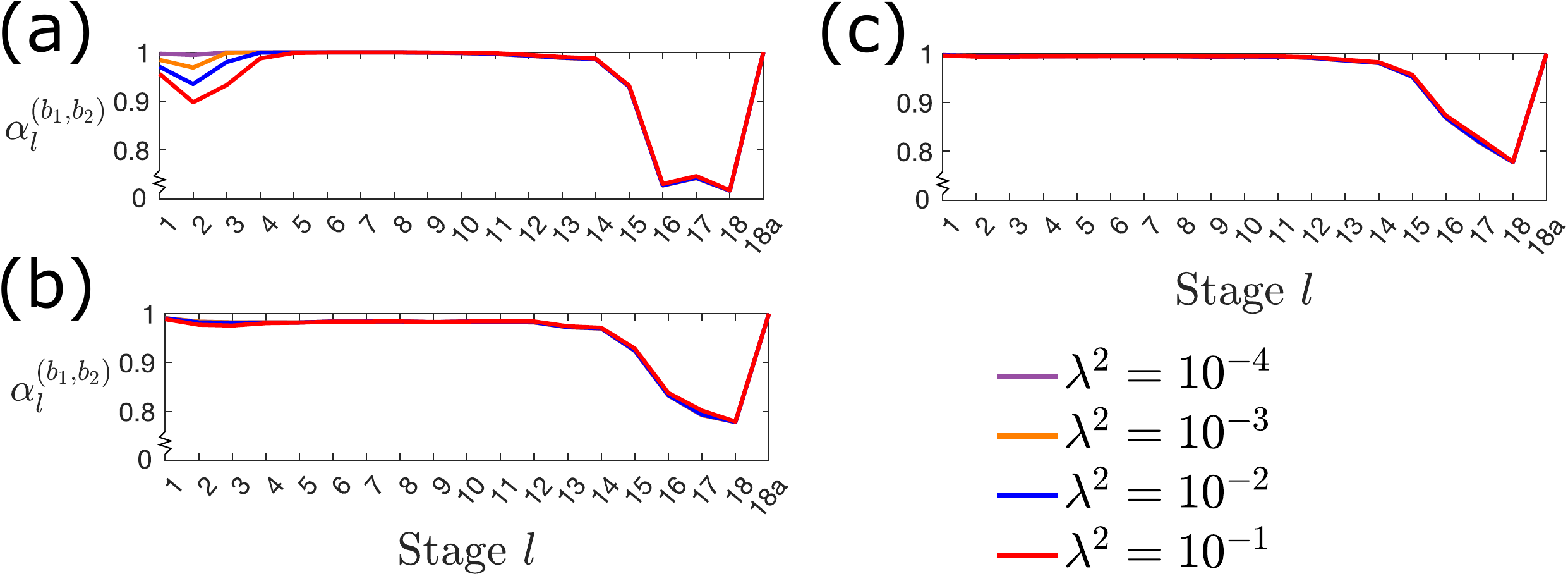}
    \caption{Robustness of DDWD to the strength of the variance penalty parameter $\lambda^2$ for synthetic signals composed of (a) Meyer, (b) fk6, and (c) db2 wavelets ($b_1$ is the corresponding wavelet basis and $b_2$ is the DDWD basis). Refer to Figure~\ref{fig:wave_compare}b for the shapes of these wavelets.}
    \label{fig:lambda_robust}
\end{figure}

First, we test the robustness of DDWD to the strength of the variance penalty parameter $\lambda^2$. To measure the ability of DDWD to uncover the wavelet from which the synthetic dataset is built, we calculate the inner product $\alpha_l^{(b_1,b_2)}$ from~\eqref{eq:alpha_fixedl} between the DDWD wavelets and the pre-specified wavelet. As in Section \ref{sec:res}, $\alpha_l^{(b_1,b_2)} \gtrsim 0.95$ is considered highly accurate. Figure~\ref{fig:lambda_robust}a shows that DDWD is accurate for stages 1 through 14 across two orders of magnitude of $\lambda^2$ when the synthetic dataset is constructed from Meyer wavelets. 

Next, we test the accuracy of DDWD when applied to synthetic datasets constructed from fk6 or db2 wavelets. Figures \ref{fig:lambda_robust}b and c show that DDWD is able to accurately recover the underlying wavelets. Furthermore, the results are extremely robust to the value of $\lambda^2$. In what follows, we fix $\lambda^2=10^{-4}$, which is the value that we use when applying DDWD to the Superpipe data. 

\begin{figure}
    \centering
    \includegraphics[width=0.8\textwidth]{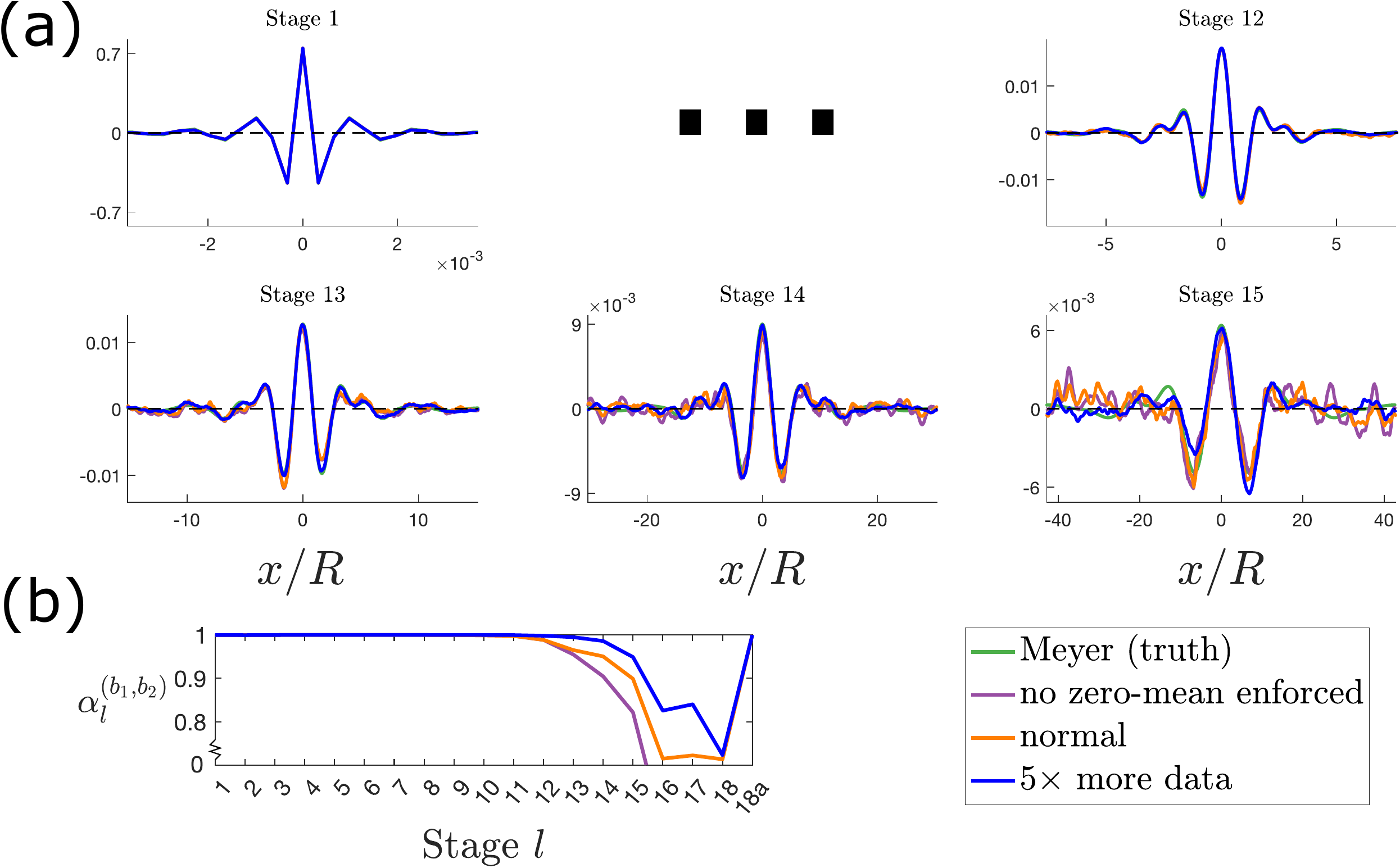}
    \caption{Results of DDWD applied to a synthetic dataset generated from Meyer wavelets using the WED of the Superpipe signal measured at the centerline. Label ``normal'' indicates that a zero mean was enforced for the DDWD wavelets. Plotted are (a) wavelets at relevant stages and (b) accuracy at each stage with $b_1$ as the true basis and $b_2$ as the computed DDWD basis.}
    \label{fig:synth_wave}
\end{figure}

Finally, we test the robustness of DDWD to the energy distribution of the synthetic signal. The synthetic dataset is constructed from Meyer wavelets and matches the WED of the Superpipe velocity at $y/R = 1$ (see Figure~\ref{fig:WED_arr}). Figure~\ref{fig:synth_wave} shows that DDWD is accurate for stages 1 through 14 when the wavelets are enforced to have zero mean (labelled ``normal''). A zero mean was enforced for the previous tests and for the application to the Superpipe data. When a zero mean is not enforced, the results are less accurate. The accuracy begins to decrease past stage 12, which is where the energy peaks. Similarly, the accuracy in Figure~\ref{fig:lambda_robust} begins to decrease past stage 14, which is where the energy peaks for the corresponding dataset. The results produced by DDWD seem to be most reliable in the range of scales where the energy monotonically increases with scale, which is likely related to DDWD minimizing energy scale by scale. As is shown, the results become more accurate as the amount of data increases. 
Overall, the validation results here indicate that DDWD is accurate for all wall-normal positions from stage 1 through stage 14.

\revise{
\subsection{Defining the length scale of a wavelet}
\label{sec:app_lambdax}		
There is no consensus for how to measure the length scale of a wavelet, which we denote with $\lambda_x$. (Note, our definition for $\lambda_x$ here is unrelated to the Fourier wavenumber $k$.) The most common approach is to calculate the center wavenumber of the wavelet, $k_c$, which is the wavenumber corresponding to either the peak or the first moment of the wavelet's PSD. However, we find that the center wavelength, $\lambda_c = 2 \pi/k_c$, is usually too small. For example, the stage-11 wavelet in Figure \ref{fig:self_sim_x}a has $\lambda_c \approx 0.6R$, which we find to not be reflective of the length scale of the wavelet since $\lambda_c$ covers approximately one oscillation while a wavelet consists of several oscillations.

Our method for measuring $\lambda_x$ involves centering an interval about the center of a wavelet and shrinking the support of the interval from both ends until it contains only 90\% of the wavelet's energy; the length of the interval is $\lambda_x$. The 90\% threshold is not arbitrary. We calculate the wavelet energy distribution (WED), which is the energy in the detail subspaces at each stage, and plot it with stage converted to $2\pi/\lambda_x$--- a 90\% threshold leads to the best overlap with the premultiplied PSD. (See more details on the WED in Section \ref{sec:energy}.) As seen with the stage-11 wavelet in Figure \ref{fig:self_sim_x}a, our method for calculating $\lambda_x=1.26 R$ seems to better reflect its length scale than $\lambda_c$.
}

\revise{
\subsection{Extending DDWD to 2D and 3D data}
While we only work with 1D data in this paper, we briefly discuss how DDWD can be extended to 2D and 3D analyses. In 1D, we maximize the energy in the approximation subspace (contains low-wavenumber information) at each stage. For each stage in 2D, we have two options for the energy maximization. The first option is to maximize the energy in the 2D subspace containing low-wavenumber information in both directions. The second option is to perform 1D DDWD in each direction, then take a tensor product of the 1D bases to get a 2D DDWD basis.
}

\bibliographystyle{jfm}
\bibliography{references}

\end{document}